\documentclass[%
floatfix,
prd,
showpacs,
nofootinbib,
superscriptaddress
]{revtex4}
\usepackage{aas_macros,amsmath,amssymb,braket,here}
\usepackage[dvipdfmx]{graphicx,color}
\input{colordvi.tex}

\usepackage{ulem}

\begin{document}
\title{Generation of magnetic fields in Einstein-Aether gravity}

\author{Shohei Saga}
\affiliation{Department of Physics and Astrophysics, Nagoya University,
Aichi 464-8602, Japan}
\email{saga.shohei@nagoya-u.jp}
\author{Maresuke Shiraishi}
\affiliation{Department of Physics and Astrophysics, Nagoya University,
Aichi 464-8602, Japan}
\author{Kiyotomo Ichiki}
\affiliation{Department of Physics and Astrophysics, Nagoya University,
Aichi 464-8602, Japan}
\affiliation{Kobayashi-Maskawa Institute for the Origin of Particles and the Universe, Nagoya University,
Nagoya 464-8602, Japan}
\author{Naoshi Sugiyama}
\affiliation{Department of Physics and Astrophysics, Nagoya University,
Aichi 464-8602, Japan}
\affiliation{Kobayashi-Maskawa Institute for the Origin of Particles and the Universe, Nagoya University,
Nagoya 464-8602, Japan}
\affiliation{Kavli Institute for the Physics and Mathematics of the Universe (Kavli IPMU), The University of Tokyo,
Chiba 277-8582, Japan}

\begin{abstract}
Recently the lower bounds of the intergalactic magnetic fields $10^{-16} \sim 10^{-20}$ Gauss are
set by gamma-ray observations while it is unlikely to generate such large scale magnetic fields through
astrophysical processes. It is known that large scale magnetic fields could be generated if
there exist cosmological vector mode perturbations in the primordial
plasma. The vector mode, however, has only a decaying solution in
General Relativity if the plasma consists of perfect fluids. In order
to investigate a possible mechanism of magnetogenesis in the primordial
plasma, here we
consider cosmological perturbations in the Einstein-Aether gravity model,
in which the aether field can act as a new source of vector metric
perturbations.
The vector metric perturbations induce the velocity difference between baryons and photons which then generate magnetic fields.
This velocity difference arises from effects at the second order in the tight-coupling approximation. We estimate the
angular power spectra of temperature and B-mode polarization of the
Cosmic Microwave Background (CMB) Anisotropies in this model and put a rough 
constraint on the aether field parameters from latest
observations. We then estimate the power spectrum of associated magnetic
fields around the recombination epoch within this limit. It is found that the spectrum
has a characteristic peak at $k=0.1 h{\rm Mpc^{-1}}$, and at that scale
the amplitude can be as large as $B\sim 10^{-22}$ Gauss where
the upper bound comes from CMB temperature anisotropies. 
The magnetic fields with this amplitude can be seeds of large scale magnetic 
fields observed today if the sufficient dynamo mechanism takes place. 
Analytic interpretation for the power spectra is also given.
 \end{abstract}
\pacs{04.50.Kd, 98.80.Es} \maketitle

\section{Introduction}
Astronomical observations have shown that magnetic fields exist
ubiquitous in various astrophysical objects, ranging from planets to clusters of
galaxies or even larger systems 
\cite{Giovannini:2003yn,Widrow:2002ud,Ryu:2011hu}. On large scales such
as in galaxies and clusters of galaxies, measurements of the Faraday
rotation indicate that the strength of magnetic fields is about $\sim
10^{-6}$ Gauss. Recently the lower bounds of the intergalactic magnetic
field are found as $10^{-16}\sim 10^{-20}$ Gauss
\cite{Neronov:1900zz,0004-637X-762-1-15}, depending on the data used and
assumptions they have made.

The origin of such large scale magnetic fields remains as an enigma and
much effort has been made to explain the origin in a wide variety of
ways. It is believed that the magnetic fields in galaxies can be
amplified and maintained by the dynamo mechanism, which is a
hydro-magnetic process with magnetic reconnections
(e.g. refs.~\cite{Widrow:2002ud,Widrow:2011hs}). However, we still need
a seed field for the dynamo process to act on. To explain the observed
magnetic fields in galaxies in the present universe, the seed field
should be as large as $10^{-20}\sim 10^{-30}$ Gauss at kpc comoving
scale~\cite{Davis:1999bt}. It is an interesting scenario that such
small seed fields may end up with the intergalactic magnetic fields 
of $\gtrsim 10^{-20}$ Gauss due to adiabatic compression.

The seed magnetic fields are possible to originate from quantum
fluctuations of the electro-magnetic fields stretched by inflation in
the early universe. If the conformal invariance is broken by some
mechanisms during the inflation era
\cite{Martin:2007ue,Subramanian:2009fu,Demozzi:2009fu,Kanno:2009ei},
magnetic fields with a large coherence length can be naturally generated
beyond the horizon scale. In this case, however, there are associated
nagging problems, namely, the back reaction and the strong coupling
problems. The former is that inflation fails to proceed if the generated
electromagnetic fields dominate the energy density in the universe
during inflation. The latter problem is related with the naturalness of
the model building. Due to these problems, there is no satisfactory
model so far to explain observed magnetic fields
\cite{Fujita:2012rb}. Nevertheless, the effects of primordial magnetic
fields on the present observations have been well investigated
\cite{Fujita:2012rb,Paoletti:2012bb,Yamazaki:2012pg,Fedeli:2012rr}.
Another possible mechanism to generate seed magnetic fields is the
phase transition in the early universe (e.g. ref.~\cite{Widrow:2011hs}).
Phase transition releases the free energy as a latent heat with forming
bubbles, and the bubble collisions can generate electric current and
hence magnetic fields. However, the magnetic fields generated at the
phase transition generally have small coherence length which
corresponds to the Hubble radius at the transition. In this case,
therefore, inverse cascading processes of magnetic fields are necessary
to explain the ones at large scales.

On the other hand, some astrophysical processes may become natural
candidates of the origin of the seed magnetic fields. For example, the Biermann
battery~\cite{1951PhRv...82..863B}, which is effective in non-adiabatic situations such as shocks, can
generate magnetic fields in starts~\cite{1982PASP...94..627K}, supernova shocks~\cite{Hanayama:2005hd,Miranda:1998ne}, large scale structure formation~\cite{Kulsrud:1996km}, and cosmological reionizatoin~\cite{Gnedin:2000ax}. However, since
the mechanism works effectively only in places where the matter or
astronomical objects are present, it may be difficult to explain the
existence of the intergalactic magnetic fields if they are in void
regions \cite{inprep}.

In this paper, we consider yet another mechanism to generate the seed
magnetic fields, i.e., the vorticity in the primordial plasma before the
recombination epoch~\cite{Harrison:1970ss}. Because photons push
electrons more frequently than ions through Thomson scatterings, the
vorticity in the photon fluid can induce circular current and thus
generate magnetic fields. The problem here is that in a Friedmann
universe with perfect fluids, the solution of the vorticity has only a
decaying mode. 
In order to have nonzero vorticity, some mechanisms are proposed, which
include the free-streaming neutrinos~\cite{Lewis:2004kg, Ichiki:2011ah},
the cosmological defects~\cite{Hollenstein:2007kg},
and the nonlinear couplings of the first order density perturbations
\cite{Ichiki:2011ah,Takahashi:2005nd,Fenu:2010kh}.

Following this idea, here we show that a possible modification of
gravity, namely the Einstein-Aether gravity~\cite{Jacobson:2000xp},
can also generate nonzero vorticity and thus magnetic fields. The
Einstein-Aether 
gravity is known as a healthy extension of Ho\v{r}ava-Lifshits gravity
at low energies~\cite{Jacobson:2010mx,Blas:2009qj}, which originally
proposed by Ho\v{r}ava~\cite{Horava:2009uw,Horava:2008ih} as a candidate
of the theory of quantum gravity. The Einstein-Aether gravity contains a new regular
vector degree of freedom which is called ``the aether field''. Effects
of the aether field have been discussed intensively,
for example, in
connection with the inflation era and late-time accelerating expansion of the universe
\cite{PhysRevD.77.084010,Zuntz:2010jp,2012PhLB..710..493M,Barrow:2012qy}, Cosmic Microwave
Background radiation (CMB) temperature anisotropies
\cite{Lim:2004js,Zuntz:2008zz,Zuntz:2010jp} and compact objects
\cite{Eling:2007xh,Garfinkle:2007bk,Eling:2006ec}.
The purpose of this paper is to examine the role of the vorticity excited by the aether
field in the generation of magnetic fields before
the recombination epoch.

The plan of this paper is as follows. In section II, we review the
Einstein-Aether gravity and describe a formalism of the vector-mode
perturbations. In section III, we focus on the evolution equation of
magnetic fields in the primordial plasma. In section IV, we explore the
evolution of the aether field and perturbations. Our main results will be
described in this section. Section V is devoted to our summary. In
Appendix A, we summarize the observational and theoretical constraints
on the aether parameters. In Appendix B, we formulate cosmological
perturbations in the Einstein-Aether gravity and define the scalar-vector-tensor decomposition.

Throughout this paper, we use the units in which $c=\hbar =1$ and the
metric 
signature as $(-, +, +, +)$. We obey the rule that the subscripts and
superscripts of the Greek characters and alphabets run from 0 to 3 and
from 1 to 3, respectively.

\section{Einstein-Aether gravity}
In this section, we summarize the Einstein-Aether gravity. 
The action for the Einstein-Aether gravity is given by~\cite{Jacobson:2000xp}
\begin{equation}
S=\frac{1}{16\pi G}\int{d^{4}x\sqrt{-g}\left[ R+\mathcal{L}_{A}\right] }+\int{d^{4}x\sqrt{-g}\mathcal{L}_{m}} \ ,
\end{equation}
where $\mathcal{L}_{A}$ is written in the form:
\begin{equation}
\mathcal{L}_{A}=K+\lambda (A^{\alpha}A_{\alpha}+1) \ , \label{Lagrangian}
\end{equation}
\begin{equation}
K\equiv K^{\alpha\beta}_{\ \ \mu\nu}\left( \nabla_{\alpha}A^{\mu}\right) \left( \nabla_{\beta}A^{\nu}\right) \ ,
\end{equation}
\begin{equation}
K^{\alpha\beta}_{\ \ \mu\nu}\equiv c_{1}g^{\alpha\beta}g_{\mu\nu}+c_{2}\delta^{\alpha}_{\ \mu}\delta^{\beta}_{\ \nu}+c_{3}\delta^{\alpha}_{\ \nu}\delta^{\beta}_{\ \mu}-c_{4}A^{\alpha}A^{\beta}g_{\mu\nu} \ .
\end{equation}
Here, $\mathcal{L}_{m}$ is the Lagrangian of the ordinary matter, $R$ is
the Ricci scalar, and $\mathcal{L}_{A}$ is the Lagrangian of the aether
field $A_{\mu}$. We assume that the aether field does not couple to the
matter. The constant $G$ is different from Newton's gravitational
constant which is locally measured. That is to say, $G$ is a ``bare''
parameter. We will see that $G$ is renormalized. The coefficients
$c_{1},\ c_{2},\ c_{3}$ and $c_{4}$ are the set of parameters of the
Einstein-Aether gravity. These parameters are constrained by
observations and theoretical hypotheses as summarized in Appendix A.
This theory contains a massive ghost~\cite{ArmendarizPicon:2009ai}, and
one usually imposes a fixed-norm constraint on the vector to remove such
a ghost. This is equivalent to forcing a Lorentz-breaking on the vacuum
expectation value. The parameter $\lambda$ is a Lagrange multiplier to
fix the norm.

To derive a set of equations of motions, we take variations with respect to variables.
Variation with respect to the Lagrange multiplier $\lambda$ imposes the condition that
\begin{equation}
A^{\mu}A_{\mu}=-1 \ . \label{Lambda} 
\end{equation}
This equation is the constraint equation of the aether field
$A^{\mu}$. Owing to the fixed norm, we will eliminate the possibility that the degree of residual freedom acts as a ghost~\cite{ArmendarizPicon:2010rs}.

Variation with respect to the aether field $A^{\mu}$ gives the equation
of motion and the continuous equation for the aether field as
\begin{equation}
\nabla_{\alpha}J^{\alpha}_{\ \ \mu}+c_{4}A^{\alpha}\left( \nabla_{\alpha}A^{\gamma}\right) \left( \nabla_{\mu}A_{\gamma}\right)=\lambda A_{\mu} \ ,\label{Aether} 
\end{equation}
where we define $J^{\alpha}_{\ \ \mu}$ as
\begin{eqnarray}\begin{split}
J^{\alpha}_{\ \ \mu}&\equiv K^{\alpha\beta}_{\ \ \mu\nu}\nabla_{\beta}A^{\nu} \\
&=c_{1}\nabla^{\alpha}A_{\mu}+c_{2}\delta^{\alpha}_{\ \mu}\nabla_{\beta}A^{\beta}+c_{3}\nabla_{\mu}A^{\alpha}-c_{4}A^{\alpha}A^{\beta}\nabla_{\beta}A_{\mu} \ .
\end{split}\end{eqnarray}

Variation with respect to the metric $g^{\mu\nu}$ gives the Einstein
equation with the aether field,
\begin{equation}
R_{\mu\nu}-\frac{1}{2}Rg_{\mu\nu}=T^{(A)}_{\mu\nu}+8\pi GT^{(M)}_{\mu\nu} \ , \label{Einstein}
\end{equation}
where we also define energy momentum tensors $T^{(A)}_{\mu\nu}$ and $T^{(M)}_{\mu\nu}$ as
\begin{eqnarray}\begin{split}
&T^{(A)}_{\mu\nu}\equiv -\frac{1}{\sqrt{-g}}\frac{\delta (\sqrt{-g}\mathcal{L}_{A})}{\delta g^{\mu\nu}} \ ,\\
&T^{(M)}_{\mu\nu}\equiv -\frac{2}{\sqrt{-g}}\frac{\delta (\sqrt{-g}\mathcal{L}_{M})}{\delta g^{\mu\nu}} \ .
\end{split}\end{eqnarray}
Note that the factors of the energy momentum tensors for the aether
field and the matter are different.
The concrete expression of $T^{(A)}_{\mu\nu}$ can be written as
\begin{equation}
T^{(A)}_{\mu\nu}=\nabla_{\sigma}\left( \mathcal{I}^{\sigma}_{\ \mu \nu}\right) +\mathcal{Y}^{(c_{1})}_{\mu\nu} -\mathcal{Z}^{(c_{4})}_{\mu\nu}+\lambda A_{\mu}A_{\nu}+\frac{1}{2}g_{\mu\nu}\mathcal{L}_{A} \ .
\end{equation}
Here we define the tensors $\mathcal{I}^{\sigma}_{\ \mu \nu}$, $\mathcal{Y}^{(c_{1})}_{\mu\nu}$ and $\mathcal{Z}^{(c_{4})}_{\mu\nu}$ as
\begin{eqnarray}\begin{split}
\mathcal{Z}^{(c_{4})}_{\mu\nu}&\equiv c_{4}A^{\alpha}A^{\beta}\left( \nabla_{\alpha}A_{\mu}\right) \left(\nabla_{\beta}A_{\nu}\right) \ , \\
\mathcal{Y}^{(c_{1})}_{\mu\nu}&\equiv c_{1}\left( \nabla^{\sigma}A_{\mu}\nabla_{\sigma}A_{\nu}-\nabla_{\mu}A^{\sigma}\nabla_{\nu}A_{\sigma}\right) \ , \\
\mathcal{I}^{\sigma}_{\ \mu\nu}&\equiv A_{(\mu}J^{\ \ \sigma}_{\nu)}-A_{(\mu}J^{\sigma}_{\ \ \nu)}-A^{\sigma}J_{(\mu\nu)} \ ,
\end{split}\end{eqnarray}
where the parentheses, ${}_{(\mu \nu)}$, mean symmetrization. For convenience we use the abbreviations, which are written in the forms:
\begin{eqnarray}\begin{split}
\alpha&= c_{1}+3c_{2}+c_{3} \ ,\\
c_{123}&=c_{1}+c_{2}+c_{3} \ ,\\
c_{13}&=c_{1}+c_{3} \ ,\\
c_{14}&=c_{1}+c_{4} \ .
\end{split}\end{eqnarray}
Because theoretical parameters have four independent components, we will treat $\alpha, c_{14}, c_{13}$ and $c_{1}$ as the independent model parameters.

\subsection{Background cosmology}
In this subsection, we explore background cosmology where the space-time
is given by a spatially flat Friedmann-Robertson-Walker metric.
The line element is given by 
\begin{equation}
ds^{2}=a^{2}(\eta)[-d\eta^{2}+d{\bf x}^{2}] \ . \label{0th metric}
\end{equation}
From Eq.~(\ref{Lambda}), the background components of the vector $A^{\mu}$
are given by
\begin{equation}
A^{\mu}=\left( a^{-1},0,0,0\right) \ . \label{0th aether}
\end{equation}
Substituting Eq.~(\ref{0th aether}) into Eq.~(\ref{Aether}), the $\mu =0$ component of
Eq.~(\ref{Aether}) can be written as
\begin{equation}
\lambda=\frac{3}{a^{2}}\left( c_{123}\mathcal{H}^{2}-c_{2}\mathcal{\dot{H}}\right) \ ,
\end{equation}
where $\mathcal{H}=\dot{a}/a$, and a dot represents a derivative
with respect to the conformal time $\eta$.
Similarly we substitute Eqs.~(\ref{0th metric}) and (\ref{0th aether}) into
Eq.~(\ref{Einstein}), and we derive the Friedmann equations with the aether field as
\begin{eqnarray}\begin{split}
\mathcal{H}^{2}&=\frac{8\pi G_{\rm cos}}{3}a^{2}\rho \ ,\\
\mathcal{\dot{H}}&=-\frac{4\pi G_{\rm cos}}{3}a^{2}(\rho +3p) \ . \label{Einstein eq}
\end{split}\end{eqnarray}
Here we define $G_{\rm cos}=\frac{G}{1-\alpha /2}$, and
$\rho$ and $p$ are the energy density and pressure of the ordinary matter, respectively. 
The effect of the aether field can only be seen in the renormalization of
Newton's gravitational constant in the background cosmology.

In the Newtonian limit which should be applied to laboratory
measurements, the Einstein equation can be rewritten in the form of the
Poisson equation~\cite{Carroll:2004ai} as
\begin{equation}
\nabla^{2}\Phi =4\pi G_{\rm N}\rho_{m} \ ,
\end{equation}
where $G_{\rm N}=\frac{G}{1+c_{14}/2}$. 
Because $G_{\rm cos}$ and $G_{\rm N}$ can be different, the ratio of $G_{\rm cos}$ and $G_{\rm N}$ is constrained by observations such as the light element abundance from the Big-Bang Nucleosynthesis (BBN)~\cite{Carroll:2004ai} (see Appendix A).
From that we have a constraint on the aether parameters as
$c_{14}+\alpha \lesssim 0.2$. We assume that $c_{14}=-\alpha$ in this
paper for simplicity, which leads to $G_{\rm cos}=G_{\rm N}$.

\subsection{Vector-mode perturbations}
In this subsection, we consider cosmological perturbations in the
Einstein-Aether gravity (see also Appendix B). Since we are interested in
generation of magnetic fields via the vector-mode perturbations, we focus
on the vector-mode in this paper, and we omit indices $^{(\pm
1)}$ or $^{(v)}$ that represent the vector-mode in Appendix B.
The scalar-vector-tensor decomposition is also defined in Appendix B.
We will work in the
synchronous gauge in which the metric is given by
\begin{equation}
ds^{2}=a^{2}\left[ -d\eta^{2}+\left( \delta_{ij}+h_{ij}\right)dx^{i}dx^{j}\right] \ ,
\end{equation}
where $h_{ij}$ is the metric perturbation.
Under this metric, the aether field can be expressed in the form as
\begin{eqnarray}\begin{split}
A^{\mu}&=\left( \frac{1}{a},\frac{V^{i}}{a}\right) \ , \\
A_{\mu}&=\left( -a,aV_{i}\right) \ ,
\end{split}\end{eqnarray}
where we have imposed $A^{\mu}A_{\mu}=-1$. Raising or lowering indices
of $V_{i}$ is done by $\delta_{ij}$. Substituting the above equations into 
Eqs.~(\ref{Einstein}) and (\ref{Aether}), we derive the evolution equations of
the perturbed quantities. The equation of motion for the aether field is given
by
\begin{equation}
c_{14}\left( \ddot{V}+2\mathcal{H}\dot{V}\right) -(\alpha -c_{14})\mathcal{\dot{H}}V+(\alpha +c_{14})\mathcal{H}^{2}V+c_{1}k^{2}V+\frac{1}{2}c_{13}k^{2}\sigma=0 \ .\label{Aether eq}
\end{equation}
The Einstein equations for the vector mode are given by
\begin{equation}
\dot{\sigma}+2\mathcal{H}\sigma=-\frac{c_{13}}{1+c_{13}}\left( \dot{V}+2\mathcal{H}V\right) -\frac{16\pi Ga^{2}p\pi}{(1+c_{13})k} \ ,
\end{equation}
\begin{equation}
k^{2}\sigma=\frac{1}{1+c_{13}}\left[ 16\pi Ga^{2}q-c_{13}k^{2}V\right] \ , \label{Constraint}
\end{equation}
where $\sigma \equiv \dot{h}^{(\pm 1)}/k$ denotes the shear and $h^{(\pm 1)}$ are the metric perturbations of the vector mode defined in Appendix~B, and $p\pi$ and $q$ denote the anisotropic stress and the heat flux of the
ordinary matter, respectively.

To see the behavior of the aether field, we solve Eq.~(\ref{Aether eq})
without the ordinary matter contribution on $\sigma$ first. The ordinary matter is subdominant
in the evolution of the vector perturbations in the
Einstein-Aether gravity, although we will include the matter
contributions correctly in our numerical calculations. This is because
the ordinary matter does not couple to the aether field while the aether field
and the metric couple directly. Thus the aether field can amplify
predominantly the metric perturbation as its source, namely, Eq.~(\ref{Constraint})
can be rewritten as
\begin{equation}
\sigma \simeq -\frac{c_{13}}{1+c_{13}}V \ .\label{re constraint}
\end{equation}
This assumption is justified by the initial conditions and the numerical
results. Using Eqs.~(\ref{Aether eq}) and (\ref{re constraint}), we can obtain
the evolution equation of the aether field in the closed form as
\begin{equation}
\ddot{V}+2\mathcal{H}\dot{V}+\left[\left( 1+\frac{\alpha}{c_{14}}\right)\mathcal{H}^{2}+\left( 1-\frac{\alpha}{c_{14}}\right)\dot{H} \right] V+c^{2}_{v}k^{2}V=0 \ ,
\end{equation}
where $c_{v}$ is the (effective) sound speed of aether field perturbation and defined by
\begin{equation}
c^{2}_{v}\equiv \frac{1}{c_{14}}\left( c_{1}-\frac{1}{2}\frac{c_{13}^{2}}{1+c_{13}}\right) \ .
\end{equation}
As a consequence of the existence of the sound speed, the aether field will
have a characteristic scale i.e., the ``sound horizon''
, which is defined as $k_{\rm SH}\equiv
aH/c_{v}$.
Depending on whether the mode is inside or outside the sound horizon, the behavior of the aether field will change. 

When we assume a long-wavelength limit $c_{v}k\eta \ll 1$ in the radiation dominated era (RD) and the matter dominated era (MD), the solutions are given by
\begin{equation}
V(k, \eta)\propto 
\begin{cases}
(k\eta)^{\nu_{\rm rad}} & {\rm RD}\\
(k\eta)^{\nu_{\rm mat}} & {\rm MD} \ , \label{long limit} \\
\end{cases}
\end{equation}
where $\nu_{\rm rad}$ and $\nu_{\rm mat}$ are defined by
\begin{eqnarray}\begin{split}
\nu_{\rm rad}&\equiv \frac{-1+\sqrt{1-8\alpha /c_{14}}}{2} \ ,\\
\nu_{\rm mat}&\equiv \frac{-3+\sqrt{1-24\alpha /c_{14}}}{2} \ .
\end{split}\end{eqnarray}
These results show that the aether field has the power-law dependence on $k\eta$ outside the sound horizon.
Because we are interested in a non-decaying regular mode in the radiation dominated era, we assume that $\alpha /c_{14}\leq 0$.
Taking into account the parameter constraint from Eq.~(\ref{Iso}), the power-law indices, $\nu_{\rm rad}$ and $\nu_{\rm mat}$, must be satisfied with the conditions $0\leq \nu_{\rm rad}\leq 1$ and $-1 \leq \nu_{\rm mat}\leq 1$.

Next, we assume the short-wavelength limit $c_{v}k\eta \gg 1$.
In this limit, the solutions are given by 
\begin{equation}
V(k, \eta)\propto 
\begin{cases}
e^{ic_{v}k\eta}/(k\eta) & {\rm RD} \\
e^{ic_{v}k\eta}/(k\eta)^{2} & {\rm MD} \ . \label{short limit} \\
\end{cases}
\end{equation}
Both in the radiation and matter dominated eras, the aether field decays as
$\propto a^{-1}$ with oscillations. 

\section{magnetic fields generation}
In this section, let us consider the generation of magnetic fields. We
follow Ref.~\cite{Ichiki:2011ah} to formulate the generation process. When there exists the
difference of velocities between baryons and photons, magnetic
fields can be generated
\cite{Ichiki:2011ah,Takahashi:2005nd,2006Sci...311..827I,2009CQGra..26m5014M,Fenu:2010kh}. This is caused by
the difference of the Thomson cross sections between electrons and
protons. In other words photons push electrons more frequently than
protons. As a consequence, the charge separation appears.
Photon's bulk pressure creates the charge separation which induces
electric fields, and then magnetic fields will be generated as well.
The evolution equation of magnetic fields is derived by the combination
of the Maxwell equations and Euler equations for electrons and
protons with Thomson scattering~\cite{Ichiki:2011ah}. It is written in
the form:
\begin{equation}
\frac{d\left( a^{2}B^{i}\right)}{dt}=\frac{4\sigma_{T}\rho_{\gamma}a}{3e}\epsilon^{ijk}\partial_{k}\left( v_{\gamma j}-v_{b j}\right) \ , \label{evolve mag1}
\end{equation}
where $\sigma_{T}$ is the Thomson cross section and $\epsilon^{ijk}$ is
the Levi-Civita symbol and $\rho_{\gamma}$ is the energy density of
photons and $v_{\gamma}$ or $v_{b}$ are the velocity of photons or baryons. When
we consider scalar perturbations for the velocity fields, 
the right-hand side of the above equation should vanish. In this paper,
we solve the above equation with an initial condition $B=0$ at $z=10^{9}$,
which roughly corresponds to the time of neutrino decoupling. By
integrating the above equation, the square of the magnetic fields is
given by
\begin{equation}
a^{4}B^{i}({\bf k},t)B^{*}_{i}({\bf k'},t)=\left( \frac{4\sigma_{T}}{3e}\right)^{2}\left( \delta^{j\ell}\delta^{km}-\delta^{jm}\delta^{k\ell}\right) k_{k}k'_{m} \int^{t}_{0}\int^{t}_{0}{a(t')\rho_{\gamma}(t')\delta v_{j}({\bf k},t')a(t'')\rho_{\gamma}(t'')\delta v^{*}_{\ell}({\bf k'},t'')dt' dt''} \ , \label{evolve mag2}
\end{equation}
where $\delta v_{j}({\bf k},t')\equiv v_{\gamma j}({\bf k},t')-v_{b j}({\bf k},t')$.
Because the evolution equations for baryons, photons and the vector-metric perturbations are independent of $\hat{k}$, we can decompose the vector-mode of the velocity difference between baryons and photons as 
\begin{equation}
\delta v({\bf k},t) = V_{\rm ini}({\bf k})\delta v (k,t) ~, \label{ini transfer}
\end{equation}
where $V_{\rm ini}({\bf k})$ is the stochastic initial amplitude of the aether field and $\delta v(k,t)$ is the transfer function of the velocity difference between baryons and photons, which is the solution with $V_{\rm ini}=1$.
In our calculations,  we will relate the initial amplitude $V_{\rm ini}$
to the initial amplitude of the aether field.

By taking an ensemble average and defining the power spectrum, we obtain
\begin{equation}
\Braket{\delta v_{j}({\bf k},t')\delta v^{*}_{\ell}({\bf k'},t'')}=(2\pi)^{3}\frac{2\pi^{2}}{k^{3}}\mathcal{P}_{V}(k)P_{j\ell}(\hat{k})\delta v(k,t')\delta v(k,t'')\delta^{3}({\bf k}-{\bf k'})\ , \label{bf ams}
\end{equation}
\begin{equation}
P_{j\ell}(\hat{k}) \equiv \delta_{j\ell}-\hat{k}_{j}\hat{k}_{\ell}\ ,
\end{equation}
where $\mathcal{P}_{V}(k)$ is the spectrum of the aether field perturbation, which is defined by
\begin{equation}
\Braket{V_{\rm ini}({\bf k})V^{*}_{\rm ini}({\bf k'})}\equiv (2\pi)^{3}\frac{2\pi^{2}}{k^{3}}{\mathcal{P}}_{V}(k)\delta({\bf k}-{\bf k'}) ~. \label{dv to aether}
\end{equation}
Because the parity is not violating in the Einstein-Aether gravity, two
helicity states $\lambda =\pm 1$ should not mix. Thus we already omit the
superscript ${}^{(\lambda)}$ here. The power spectrum of magnetic fields is defined as
\begin{equation}
\Braket{B^{i}({\bf k},t)B^{*}_{i}({\bf k'},t)} \equiv (2\pi)^{3}S_{B}(k,t)\delta^{3}({\bf k}-{\bf k'})\ .
\end{equation}
Then we can express $S_{B}(k,t)$ by the initial power spectrum for the aether field $\mathcal{P}_{V}(k)$ as
\begin{equation}
a^{4}(t)\frac{k^{3}}{2\pi^{2}}S_{B}(k,t)=\left( \frac{4\sigma_{T}}{3e}\right)^{2} 2\mathcal{P}_{V}(k)k^{2}\left[ \int^{t}_{0}{dt' a(t')\rho_{\gamma}(t')\delta v(k,t')}\right]^{2} \ . \label{Mag}
\end{equation}
The initial power spectrum may depend on the inflation model
considered. Generally, the initial power spectrum is assumed to be given by a power law as
\begin{equation}
\mathcal{P}_{V}(k)=\mathcal{A}_{V}\left( \frac{k}{k_{0}}\right)^{n_{v}} \ , \label{Initial power}
\end{equation}
where $k_{0}=0.002{\rm Mpc^{-1}}$ .
If we fix the inflation model and the subsequent reheating model, the initial amplitude
$\mathcal{A}_{V}$ and spectral index $n_{v}$ can be expressed as a
function of the aether parameters and inflation model parameters (see Appendix B).

The amplitude of velocity difference between baryons and photons to generate
magnetic fields is dependent on the amplitude of the metric perturbation of the
vector-mode. The aether field induces the metric perturbation of the
vector-mode dominantly, and therefore the evolution of the aether field
is closely related to that of the magnetic fields and their power
spectrum.

To understand the behavior of the vector-mode velocities in the
presence of aether field, we solve the Euler equation of the baryons for vector-mode.
The vector-mode evolution equation for baryons is given by~\cite{Lewis:2004kg},
\begin{equation}
\dot{v}_{b}+\mathcal{H}v_{b}=-\frac{4\rho_{\gamma}}{3\rho_{b}}an_{e}\sigma_{T}(v_{b}-v_{\gamma}) \ , \label{evo vb}
\end{equation}
where $n_{e}$ is the electron number density. 
For photons, we perform a multipole expansion of the Boltzmann equation of photons for the vector-mode as~\cite{Lewis:2004kg}
\begin{eqnarray}\begin{split}
&\dot{v}_{\gamma}+\frac{1}{8}k\pi_{\gamma}=-an_{e}\sigma_{T}(v_{\gamma}-v_{b}) \ , \\
&\dot{\pi}_{\gamma}+\frac{8}{5}kI_{3}-\frac{8}{5}kv_{\gamma}=-an_{e}\sigma_{T}\left( \frac{9}{10}\pi_{\gamma}-\frac{9}{5}E_{2}\right) -\frac{8}{5}k\sigma \ , \\
&\dot{I}_{\ell}+k\frac{\ell}{2\ell +1}\left( \frac{\ell +2}{\ell +1}I_{\ell +1} -I_{\ell -1}\right) =-an_{e}\sigma_{T}I_{\ell} \ , \label{Boltzmann eq}
\end{split}\end{eqnarray}
where $I_{\ell}$ is the $\ell$th order moment of the photon distribution function and $E_{\ell}$ is the $\ell$th order moment of E-mode polarization.
The aether field does not affect the Boltzmann equation for photons and the Euler equation for baryons.
However the aether field may change the initial conditions for the matter part because the Einstein equation contains the aether field.

Before moving to our numerical calculations, we summarize the initial conditions and the set of equations under the tight-coupling approximation.
The initial conditions with the aether field have been derived by
assuming that the universe is deep in the radiation dominated era and
expanding the equations in powers of $k\eta$ up to the lowest order
\cite{Nakashima:2011fu}.
Because we are interested in the effects of the aether field, we ignore the regular vector-mode in the presence of the neutrino anisotropic stress which is investigated in~\cite{Ichiki:2011ah}.
Then the aether field in powers of $k\eta$ up to the lowest order is given by
\begin{equation}
V=V_{\rm ini}\eta^{\nu_{\rm rad}}\left[1-\left( 1-\frac{\nu_{\rm rad}}{2}\right)\frac{\omega \eta}{4} \right] \ ,
\end{equation}
where $\omega =\Omega_{m}\mathcal{H}_{0}/\sqrt{\Omega_{R}}$ and $\Omega_{i}$ are ordinary cosmological density parameters.
The initial conditions for the other variables are given by
\begin{eqnarray}\begin{split}
v_{\gamma}&=0 \ , \\
v_{\nu}&=0 \ , \\
\sigma &= -\frac{\nu^{*}_{\rm rad}}{\nu^{*}_{\rm rad}+4R^{*}_{\nu}}\frac{c_{13}}{1+c_{13}}V_{\rm ini}\eta^{\nu_{\rm rad}} \ , \\
\frac{\pi_{\nu}}{\rho_{\nu}}&=-\frac{8}{15(1+\nu_{\rm rad})}\frac{\nu^{*}_{\rm rad}}{\nu^{*}_{\rm rad}+4R^{*}_{\nu}}\frac{c_{13}}{1+c_{13}}V_{\rm ini}k\eta^{1+\nu_{\rm rad}} \ ,
\end{split}\end{eqnarray}
where $R^{*}_{\nu}=(1-\alpha /2)/(1+c_{13}) R_{\nu}$, $R_{\nu}=\Omega_{\nu}/\Omega_{r}$, $\nu^{*}_{\rm rad}=5(1+\nu_{\rm rad})(2+\nu_{\rm rad})/2$.
We apply these initial conditions to our numerical calculations.
In our numerical calculation, we set $V_{\rm ini}=1$ and
multiply the power spectrum $P_{V}(k)$ when we derive the variance of the perturbation variables as in Eq.~(\ref{dv to aether}).

Deep in the radiation dominated era, photons and baryons frequently interact with each other.
These fluids are tightly coupled because the opacity $\dot{\tau}=an_{e}\sigma_{T}$ is large.
Hence the tight-coupling parameter $\epsilon$ allows us to expand the equations.
The tight-coupling parameter can be expressed as
\begin{equation}
\epsilon =\frac{k}{\dot{\tau}}\sim 10^{-2}\left( \frac{k}{1{\rm Mpc^{-1}}}\right)\left( \frac{1+z}{10^{4}}\right)^{-2}\left( \frac{\Omega_{b}h^{2}}{0.02}\right)^{-1} \ ,\label{Epsilon}
\end{equation}
where $\Omega_{b}$ is the baryon density normalized by the critical
density and $h\equiv H_{0}/100\; (\rm{km\; s^{-1}\; Mpc^{-1}})$ is the
normalized Hubble parameter with $H_{0}$ being the Hubble constant. We
expand the equations using the tight-coupling parameter up to second
order~\cite{Ichiki:2011ah}. From here, we quickly review the derivation of expanded equations using the tight-coupling parameter up to the second order.
At the zeroth order, Eqs.~(\ref{evo vb}) and (\ref{Boltzmann eq}) give
\begin{eqnarray}\begin{split}
v^{(0)}_{\gamma}-v^{(0)}_{b}&=0 ~,\\
\pi^{(0)}_{\gamma}=2E^{(0)}_{2}&=0 ~, \\
\dot{v}^{(0)}_{\gamma}+\frac{R\mathcal{H}}{1+R}v^{(0)}_{\gamma}&=0 ~,
\end{split}\end{eqnarray}
where $R = \frac{3\rho_{b}}{4\rho_{\gamma}}$ and the superscript means the order of the tight-coupling parameter.
At the first order, Eq.~(\ref{evo vb}) and the first line of Eq.~(\ref{Boltzmann eq}) give
\begin{equation}
v^{(1)}_{\gamma}-v^{(1)}_{b}=\frac{k}{\dot{\tau}}\left[ \frac{1}{k}\frac{R\mathcal{H}}{1+R}v^{(0)}_{\gamma}\right] ~. \label{1st order dv}
\end{equation}
From the second line of Eq.~(\ref{Boltzmann eq}), we derive the anisotropic stress of the photons up to the first order as
\begin{equation}
\pi^{(1)}_{\gamma}=\frac{k}{\dot{\tau}}\left[ \frac{32}{15}\left( v^{(0)}_{\gamma}+\sigma^{(0)}\right) \right] ~. \label{1st order pi}
\end{equation}
Note that the anisotropic stress of the photons is sourced by the shear and generated far into the radiation dominated era.
Finally, to derive the tight-coupling solution of the velocity
difference at the second order, we use Eq.~(\ref{evo vb}), the first line of Eq.~(\ref{Boltzmann eq}) and also Eqs.~(\ref{1st order dv}) and (\ref{1st order pi}) and obtain
\begin{equation}
v^{(2)}_{\gamma}-v^{(2)}_{b}=\frac{k}{\dot{\tau}}\left[ \frac{R\mathcal{H}}{(1+R)k}v^{(1)}_{\gamma}\right]-\frac{4}{15}\left( \frac{k}{\dot{\tau}}\right)^{2}\left[ \frac{R}{1+R}(v^{(0)}_{\gamma}+\sigma^{(0)})\right] ~, \label{Tight coupling}
\end{equation}
where we have neglected the cosmological redshift terms.
Therefore, up to the second order in the tight-coupling approximation, the velocity difference between baryons and photons is given as
\begin{equation}
\delta v=\frac{k}{\dot{\tau}}\frac{R\mathcal{H}}{(1+R)k}v_{\gamma}-\frac{4}{15}\left( \frac{k}{\dot{\tau}}\right)^{2}\frac{R}{1+R}(v_{\gamma}+\sigma) ~. \label{Tight coupling}
\end{equation}
At the first order in the tight-coupling approximation, the shear does not contribute to the velocity difference between baryons and photons  $\delta v$ as shown in Eq.~(\ref{1st order dv}).
However if the equations are expanded up to the second order, $\delta v$ is sourced from the shear because $\pi^{(1)}_{\gamma}$ is also sourced from the shear as Eq.~(\ref{1st order pi}).

In Eq.~(\ref{Tight coupling}), although the first term does not depend on the wavenumber, the second term depends on $k^{2}$.
Because the initial conditions of our numerical calculations imply that $\sigma \gg v_{\gamma}$, we can see that the second term dominates over the first term.
In contrast, if we consider the vector-mode without the aether field, $v_{\gamma}$ and $\sigma$ are in the same order.
In this case, the first term in Eq.~(\ref{Tight coupling}) must dominate in the early universe~\cite{Ichiki:2011ah}.
This fact appears in the difference of $k$ dependence of $\delta v$ as shown later.
Namely, in the case without the aether field, we have asymptotic scaling as $\delta v \propto k^{0}$.
In contrast, in the case with the aether field, the scaling relation becomes $\delta v\propto k^{2}$.
\section{Numerical results}
\subsection{Time evolution}
In this section, we show the result of our numerical calculations.
To solve the set of equations, namely, the evolution
equation of the aether field (Eq.~(\ref{Aether})), the Einstein equation with the
aether field (Eq.~(\ref{Einstein})) and the Boltzmann equation for the photon distribution function (Eq.~(\ref{Boltzmann eq})), we modified the CAMB code~\cite{Lewis:1999bs}.

\begin{figure}[htp]
\begin{minipage}[m]{0.45\textwidth}
\rotatebox{0}{\includegraphics[width=1.0\textwidth]{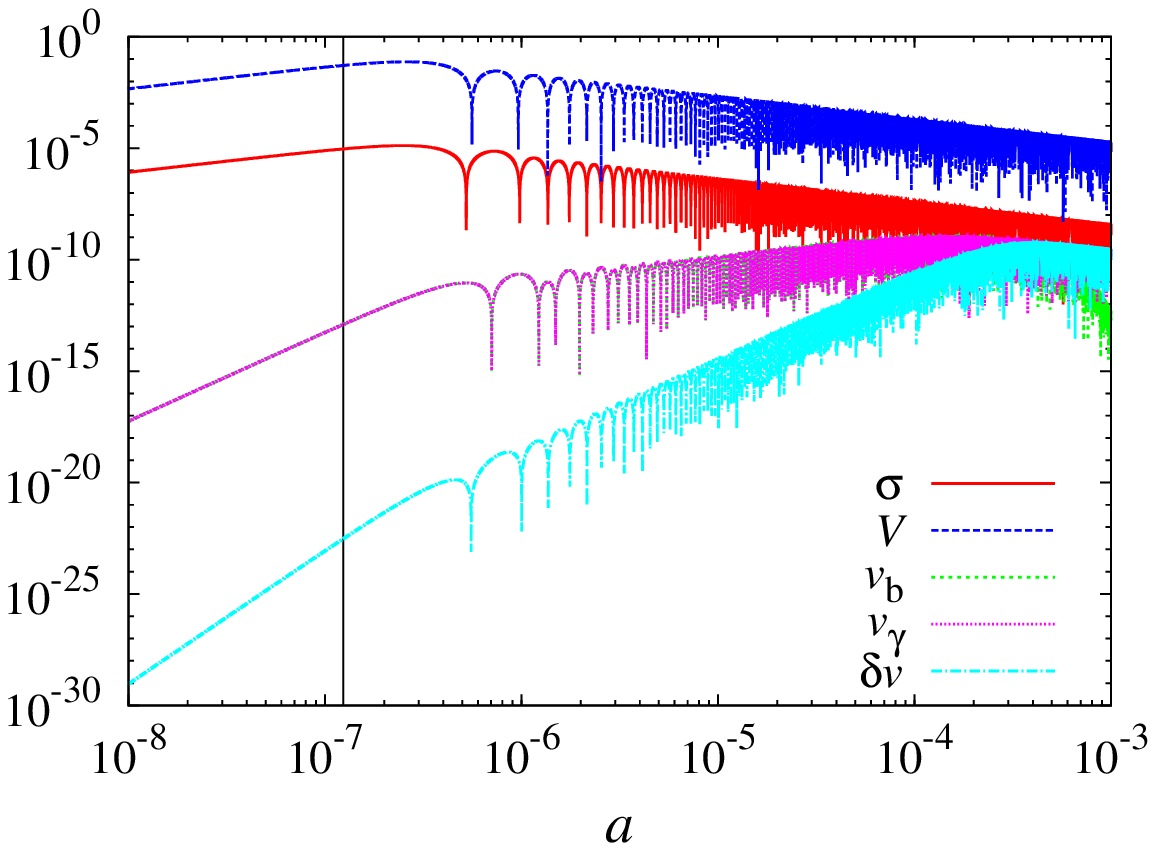}}
\end{minipage}
\begin{minipage}[m]{0.45\textwidth}
\rotatebox{0}{\includegraphics[width=1.0\textwidth]{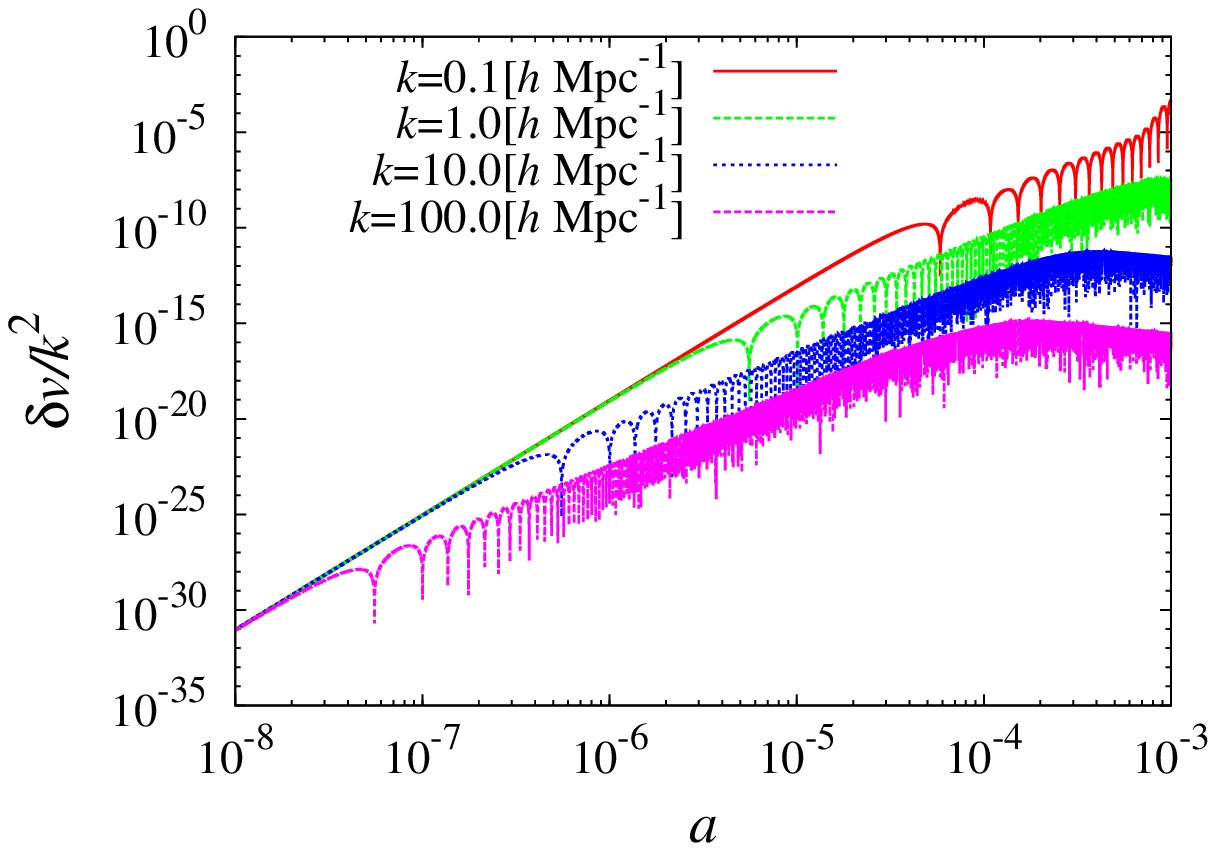}}
\end{minipage}
\caption{Time evolutions of perturbed variables in the Einstein-Aether
gravity for the aether parameters fixed as Eq.~(\ref{Params1}) at $k=10\:
h{\rm Mpc^{-1}}$. Shown in the left panel are the shear $\sigma$,
the aether field $V$, the velocity of baryons $v_{b}$, the velocity of
photons $v_{\gamma}$ and the velocity difference between baryons and
photons $\delta v$. Black solid vertical line does not express a
horizon entry time but ``a sound horizon entry time''. The metric
perturbation of the vector-mode $\sigma$ and the aether field start to
decay not after the horizon crossing but after
the sound horizon crossing. Subsequently, they start to oscillate and
damp. In the right panel time evolutions of the velocity difference
between baryons and photons $\delta v$ are shown for some
wavenumbers as indicated. We normalized by the factor of
 $k^{2}$ to remove the
scale dependence in early time (see the last
paragraph of the previous
section). } \label{fig:evolve12}
\end{figure}

In order to see the evolution of perturbed quantities, we show two results of each parameter set, and the magnitude of these parameter sets is extremely different from each other.
First, we fix the aether parameters as
\begin{eqnarray}\begin{split}
c_{14}&=-1.0\times 10^{-4} \ ,\\
c_{13}&=-2.0\times 10^{-4} \ ,\\
\alpha =-c_{14}&=1.0\times 10^{-4} \ ,\\
c_{1}&=-3.0\times 10^{-4} \ ,\\
c^{2}_{v}&=3.0 \ . \label{Params1}
\end{split}\end{eqnarray}
These values are consistent with all the constraints coming from the observations and the theoretical hypotheses as shown in Appendix A.
Time evolutions of perturbed variables are depicted in Fig.~\ref{fig:evolve12}. 
As mentioned earlier, the perturbation of the aether field $V$ and the metric perturbation
$\sigma$ dominate over the other perturbation variables in the early
universe. When the modes of $\delta v$ and $\sigma$ enter the sound
horizon, they start to oscillate and decay in proportion to the inverse
of the scale factor. On the contrary the baryon and the photon velocities
increase until breaking the tight-coupling.

When the Fourier modes reach the Silk damping scale, perturbations in
the photon fluid start to vanish exponentially along with the baryon
velocity and the potential $\sigma$ in the absence of the aether field
\cite{Ichiki:2011ah}. In contrast, in the presence of the aether field, such
exponential damping does not appear as shown in Fig.~\ref{fig:evolve12}.
This is because the aether field can keep the vector potential $\sigma$ large
enough, and push the photon fluid in equilibrium with $\sigma$ through
the photon anisotropic stress even in the photon diffusion regime. The
baryon velocity decays faster than the photon velocity in a particular case shown in Fig.~\ref{fig:evolve12} (green dashed line).
This can be understood by noting the fact that baryons are non-relativistic having no anisotropic stress,
and therefore their velocity is completely determined
by the Compton drag force from the photon fluid independently of the metric perturbation $\sigma$ (see Eq.~(\ref{evo vb})).
Because the photon velocity is sustained by the aether field and kept large enough
even after the tight-coupling is broken down, the equation for the baryon velocity Eq.~(\ref{evo vb}) is reduced to
\begin{equation}
\dot{v}_{b}\approx \frac{4\rho_{\gamma}}{3\rho_{b}}an_{e}\sigma_{T}v_{\gamma} \ .
\end{equation}
Because the right hand side of the above equation starts to damp exponentially
with rapid oscillations right before the recombination epoch, the baryon velocity also damps exponentially with oscillations.
As a result, the difference of the velocities between baryons and photons has the same
amplitude as the photon velocity at late time.

In Fig.~\ref{fig:evolve12}, before crossing the sound horizon, the
evolution of $\delta v$ may be explained by use of the tight-coupling
approximations, i.e., Eqs.~(\ref{Constraint}), (\ref{Epsilon}) and (\ref{Tight coupling}).
Since $\sigma$ and the aether field dominate the perturbation variables, we can ignore the photon velocity in Eq.~(\ref{Tight coupling}).
Then we have an approximate expression as
\begin{equation}
\delta v=-\frac{4}{15}\left( \frac{k}{\dot{\tau}}\right)^{2}
 \frac{R}{1+R}\sigma \ .
\label{approx expression}
\end{equation}
Substituting the time dependence of $\dot{\tau}$, $R$ and $\sigma$ during
the radiation dominated era, the above equation is reduced to the simple form as
$\delta v \propto k^{2}a^{5}\sigma \propto k^{2}a^{5+\nu_{\rm rad}}$.
After crossing the sound horizon, $\delta v$ continues to grow until
the tight-coupling is broken. Subsequently, the velocity difference 
between baryons and photons $\delta v$ decays together with $\sigma$
as discussed above.

\begin{figure}[htp]
\begin{minipage}[m]{0.45\textwidth}
\rotatebox{0}{\includegraphics[width=1.0\textwidth]{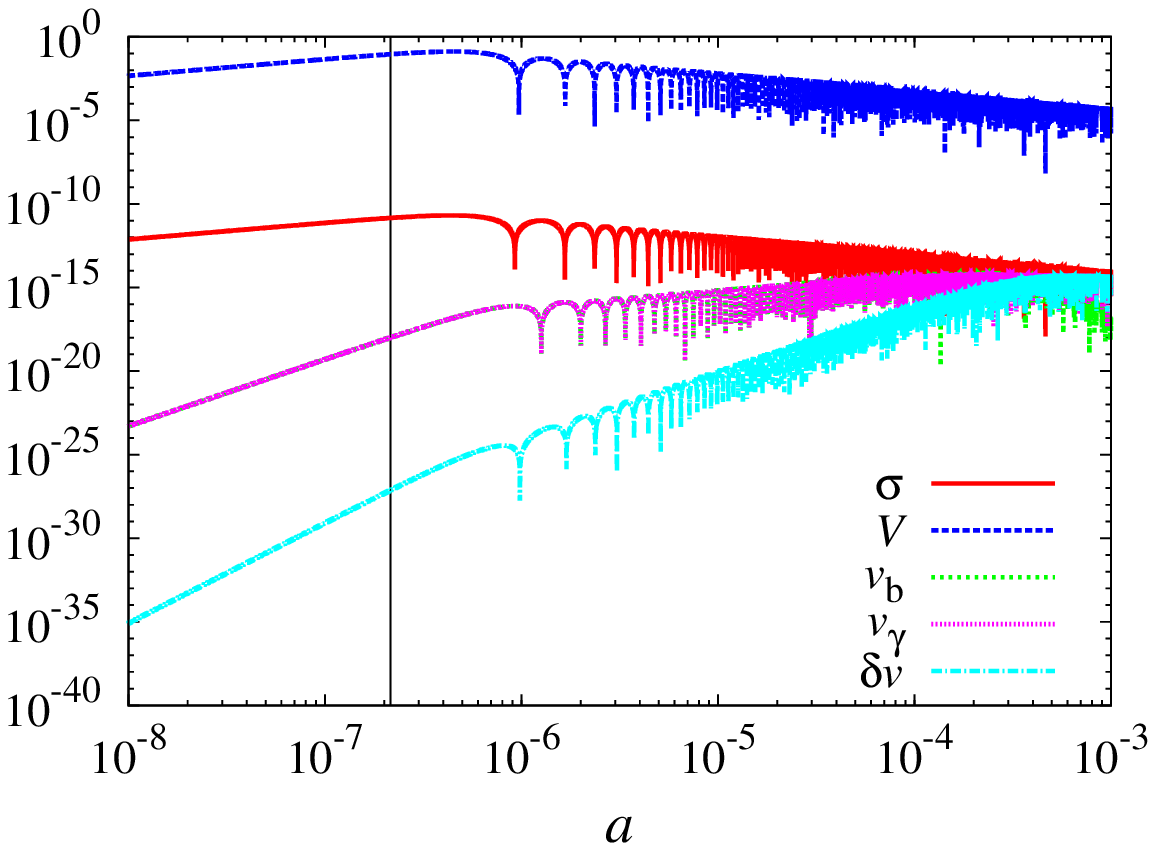}}
\end{minipage}
\begin{minipage}[m]{0.45\textwidth}
\rotatebox{0}{\includegraphics[width=1.0\textwidth]{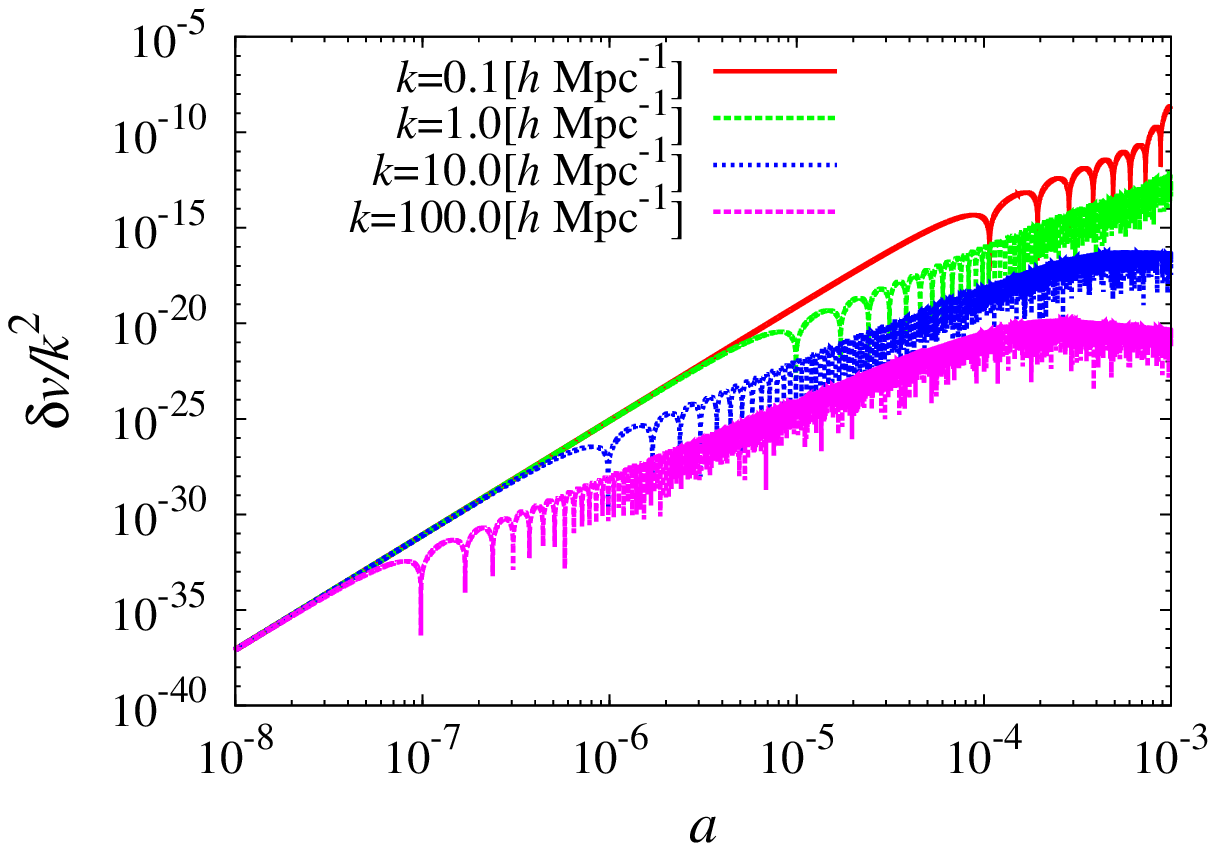}}
\end{minipage}
\caption{Same as Fig.~\ref{fig:evolve12} but for the different aether parameters fixed as Eq.~(\ref{Params2}).}
\label{fig:evolve45}
\end{figure}

To see the dependence of the aether parameters, we consider another
parameter set:
\begin{eqnarray}\begin{split}
c_{14}&=-0.9\times 10^{-10} \ ,\\
c_{13}&=-1.8\times 10^{-10} \ ,\\
\alpha =-c_{14}&=0.9\times 10^{-10} \ ,\\
c_{1}&=-0.9\times 10^{-10} \ ,\\
c^{2}_{v}&=1.0 \ . \label{Params2}
\end{split}\end{eqnarray}
The result is depicted in Fig.~\ref{fig:evolve12} and Fig.~\ref{fig:evolve45}. 
The relation between $V$ and $\sigma$ obtained from Eq.~(\ref{re constraint}) is indeed satisfied.
We find that $\delta v$ in Fig.~\ref{fig:evolve45} is smaller than one in Fig.~\ref{fig:evolve12}.
If $c_{13} \ll 1$, we can rewrite the equation (\ref{re constraint}) as $\sigma \simeq -c_{13}V$.
Because of this relation, $\sigma$ is suppressed by the factor of
$c_{13}$ compared with $V$. 
Furthermore, we can see $\delta v \sim v_{\gamma}\sim \sigma$ at late time.
Therefore, magnetic fields are also suppressed at late time by the
factor of $c_{13}$ if $c_{13}\ll 1$.

\subsection{Power spectra of temperature and B-mode polarization}
In this subsection, we calculate the power spectra of CMB temperature and
B-mode polarization. We assume that the initial power spectrum of the
aether field is given by a power law as Eq.~(\ref{Initial power}).
Although the initial amplitude $\mathcal{A}_{V}$ and spectral index $n_{v}$ may depend
on the early universe model, following the discussion in Ref.~\cite{Nakashima:2011fu}, we assume that
\begin{equation}
n_{v}=3-\sqrt{1-\frac{\alpha}{c_{14}}\frac{4\varepsilon}{(1-\varepsilon)^{2}}} \ ,
\end{equation}
where $\varepsilon =1-\mathcal{H}'/\mathcal{H}^{2}$ is the slow-roll
parameter. In addition, we assume that the spectral index does not
change in the scales of interest. The combination of the
aether parameters $\alpha /c_{14}$ is constrained as $-1\leq \alpha
/c_{14}\leq 0$ for perturbation of the aether field
to have a growing mode solution in the radiation 
dominated era and the isocurvature mode does 
not grow, see Eq.~(\ref{Iso}). The spectral index is also constrained as
$2-2\varepsilon/(1-\varepsilon)^{2}\lesssim n_{v}\leq 2$ for $\varepsilon \ll 1$, 
and thus we fix the spectral index as $n_{v}=2$ in our
numerical calculations for simplicity.
Since the aether amplitude may further change during the reheating stage,
we treat the initial amplitude as a free parameter~\cite{Nakashima:2011fu}. 

\begin{figure}[htp]
\begin{minipage}[m]{0.45\textwidth}
\rotatebox{0}{\includegraphics[width=1.0\textwidth]{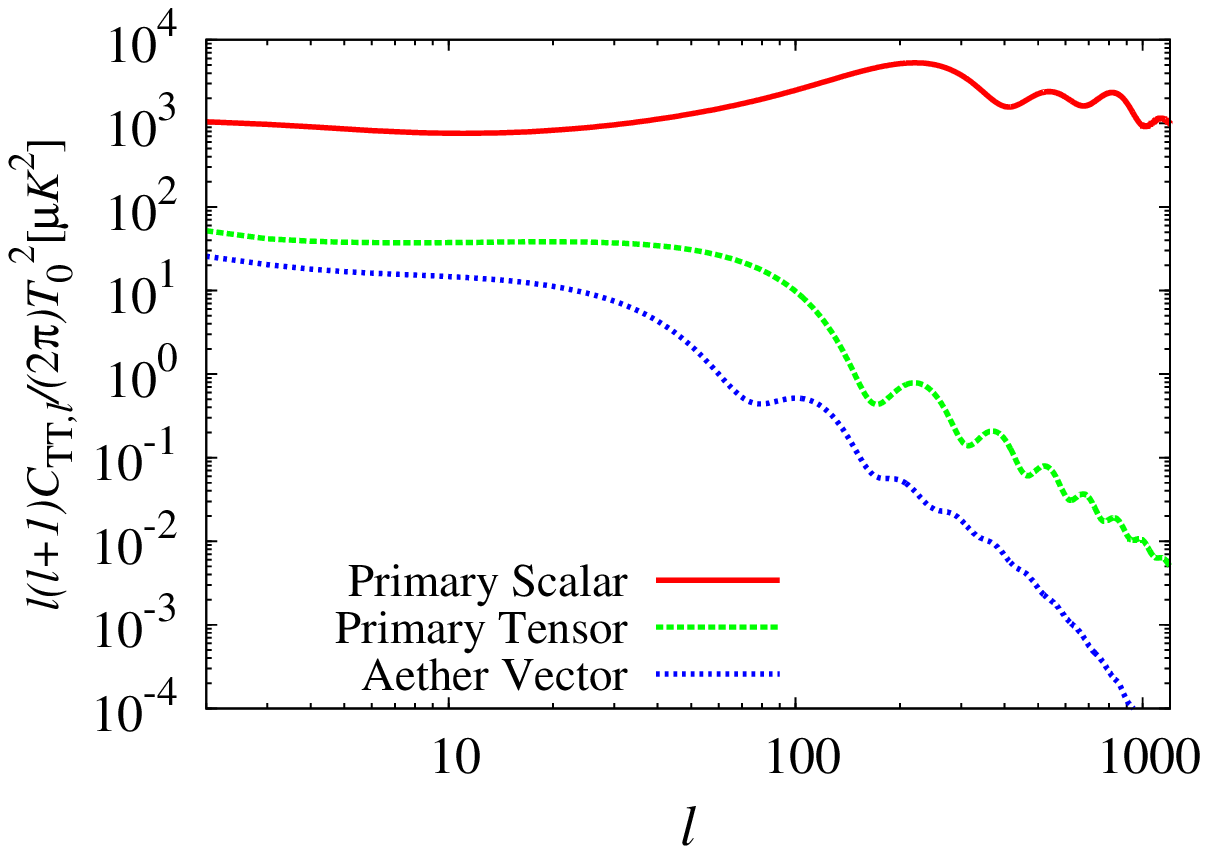}}
\end{minipage}
\begin{minipage}[m]{0.45\textwidth}
\rotatebox{0}{\includegraphics[width=1.0\textwidth]{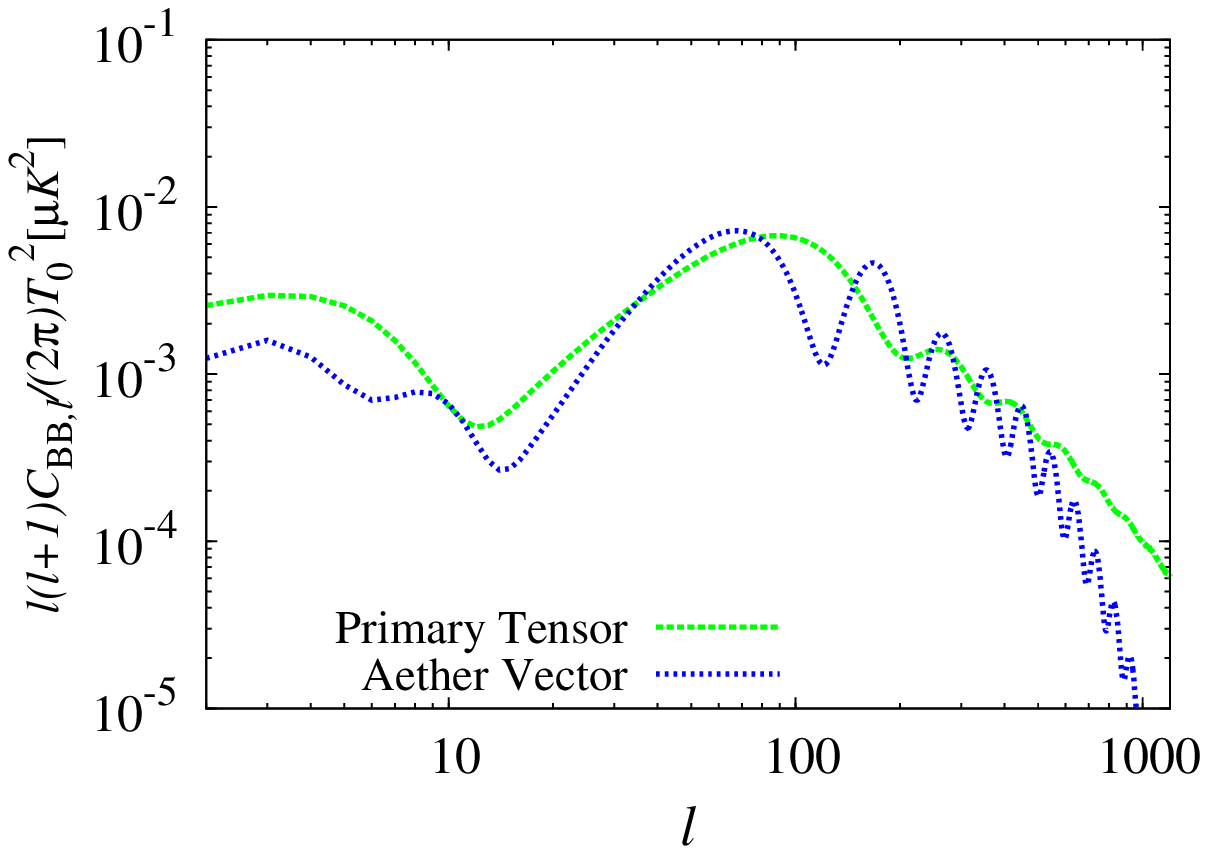}}
\end{minipage}
\caption{
The CMB temperature and B-mode polarization anisotropy power spectra.
``Primary Scalar'' and ``Primary Tensor'' indicate the primary spectra from the scalar and tensor perturbations in the standard cosmology.
``Aether Vector'' indicates the spectra from the vector mode perturbation in the Einstein-Aether gravity.
The aether parameters are set as $c_{14}=-1.0\times 10^{-4}$, $c_{13}=-2.0\times 10^{-4}$, $\alpha=1.0\times 10^{-4}$, $c_{1}=-3.0\times 10^{-2}$ and $c^{2}_{v}=3.0$.
The initial amplitude and the spectral index are fixed as $\mathcal{A}_{V}c^{2}_{13}c^{-4}_{v}=2.0\times 10^{-17}$, and
$n_{v}=2.0$ respectively.
We assume that the tensor-to-scalar ratio is $r=0.1$. 
}
\label{fig:CMB1}
\end{figure}

\begin{figure}[htp]
\rotatebox{0}{\includegraphics[width=0.5\textwidth]{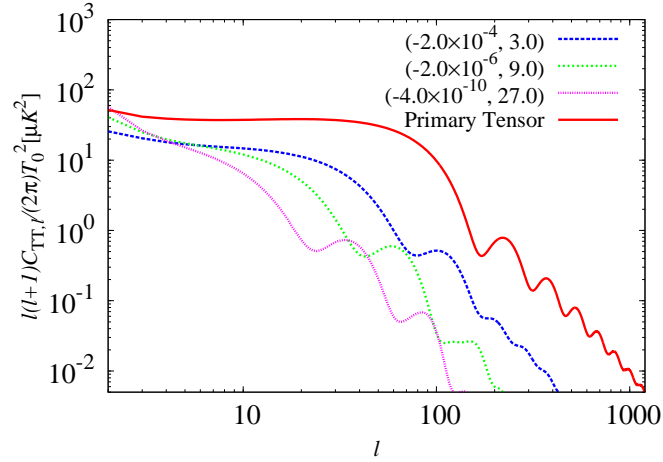}}
\caption{
The CMB temperature anisotropy (TT) power spectra with different aether parameters; $c_{13}$ and $c_{v}$.
Here we impose the condition, $\mathcal{A}_{V}c^{2}_{13}c^{-4}_{v}=2.0\times 10^{-17}$,
on the aether parameters and the initial amplitude.
``Primary Tensor'' indicate the primary spectrum from the tensor perturbation in the standard cosmology.
The spectra from the aether perturbation are depicted by solid lines,
$(c_{13},\ c^{2}_{v})=$
$(-2.0\times 10^{-4},\ 3.0)$,
$(-2.0\times 10^{-6},\ 9.0)$ and
$(-4.0\times 10^{-10},\ 27.0)$, respectively.
This figure suggests that if $\mathcal{A}_{V}c^{2}_{13}c^{-4}_{v}$ is fixed,
the amplitudes of TT power spectra are almost the same at low multipoles.
On the contrary, the peak locations are shifted.
The TT power spectrum from tensor modes with $r=0.1$ is also plotted as
 a reference (broken line). 
}
\label{fig:CMB compare}
\end{figure}

Before moving to the calculation of the magnetic field spectrum,
we can give a rough constraint on the initial amplitude and the aether
parameters from the CMB temperature anisotropies. The CMB temperature
(TT) and B-mode polarization (BB) power spectra for the fiducial
parameter set are depicted in Fig.~\ref{fig:CMB1}. From the figure we find
that the TT power spectrum from the aether field looks similar to that
from the primary tensor perturbations, with a slightly different horizon
scale, i.e., the first peak location, as discussed in section
IV. A. Therefore the constraint on the aether field from the TT power
spectrum comes from the low multipole components, which is similar to the case of 
the primary tensor modes. As for the parameter dependence of the
TT and BB power spectra, we find that the amplitudes of the spectra depend on the
combination of the initial amplitude ${\cal A}_{V}$, the aether
parameters $c_{13}$ and $c_{v}$ as ${\cal A}_{V}
c_{13}^2c^{-4}_{v}$. One can understand this dependence as follows.
First of all, the amplitude of
the metric perturbation $\sigma$ is proportional to $c_{13}$ as shown in
Eq.~(\ref{re constraint}), and thus $C_\ell\propto \sigma^2
\propto c_{13}^2$. 
Next, 
from Eqs.~(\ref{long limit}) and (\ref{short limit}), the aether
perturbation $V$ evolves as $ V \propto \eta$ outside the sound horizon
during the radiation dominated era while $V$ decays as $V \propto
\eta^{-1}$ with oscillations inside the sound horizon. 
The resultant amplitude inside the sound horizon can be expressed as $V = V_{\rm
ini}\left({\eta_{\ast}}/{\eta_{\rm
ini}}\right)\left({\eta}/{\eta_{\ast}}\right)^{-1}$
where the subscript ``ini'' means the initial value and $\eta_{\ast}$
represents the epoch of sound horizon crossing, which is given by
$\eta_{\ast} = c_v^{-1}$. 
Therefore $V \propto \eta_{\ast}^2 = c_v^{-2}$ and the amplitude of the
spectra will be changed in proportion to $c^{-4}_{v}$. Note that the
peak positions of the spectra also depend on $c_{v}$.
We show this dependence explicitly in Fig.~\ref{fig:CMB compare}.

From the above discussion and Fig.~\ref{fig:CMB1},
we find a new constraint on the aether parameters and the initial amplitude.
Because the dependence of $C^{\rm aether}_{\ell}$ on the aether parameters is proportional to
$\mathcal{A}_{V}c^{2}_{13}c^{-4}_{v}$, we find a relation from Fig.~\ref{fig:CMB1} as
\begin{equation}
C^{\rm aether}_{\ell \ TT} \approx C^{{\rm GW}(r=0.1)}_{\ell\ TT} \left(
 \frac{\mathcal{A}_{V}c^{2}_{13}c^{-4}_{v}}{2.0\times 10^{-17}}\right)
\ ~,
\end{equation}
where $C^{{\rm GW}(r=0.1)}_{\ell\ TT}$ is the
TT power spectrum by primordial gravitational waves with the scalar-tensor ratio $r=0.1$.
By imposing $C^{\rm aether}_{\ell} \lesssim
C^{{\rm GW}(r=0.1)}_{\ell\ TT}$, which is the current upper bound
obtained by the WMAP team~\cite{Bennett:2012fp,Hinshaw:2012fq},
we can place a constraint on the aether parameters and the initial amplitude as 
\begin{equation}
\mathcal{A}_{V}c^{2}_{13}c^{-4}_{v}\lesssim 2.0\times 10^{-17} \ . \label{new constraint}
\end{equation}
When we assume the single field slow-roll inflation, the initial amplitude $\mathcal{A}_{V}$ is given in the aether parameters (See Appendix B).
If we adopt Eq.~(\ref{ini amp}), the above inequality can be rewritten as $c^{2}_{13}(-c_{14})^{-1}c^{-5}_{v}\lesssim 10^{43}$.

It should be noted that the B-mode power spectrum
shows distinctive feature of oscillations at $\ell \gtrsim 80$, and thus
future precise 
measurements of B-mode polarizations can be used to discriminate the
contributions from the aether field and the primary tensor
perturbations (Ref.~\cite{Nakashima:2011fu}). Note also that the aether field,
with this upper bound on ${\cal A}_{V} c_{13}^{2}c^{-4}_{v}$, 
gives negligible contributions to the TE (E-mode and temperature cross
correlation) and EE (E-mode auto correlation) power spectra compared
to those from the standard (observed) density perturbations, and
we have omitted them here.
\begin{figure}[htp]
\rotatebox{0}{\includegraphics[width=0.5\textwidth]{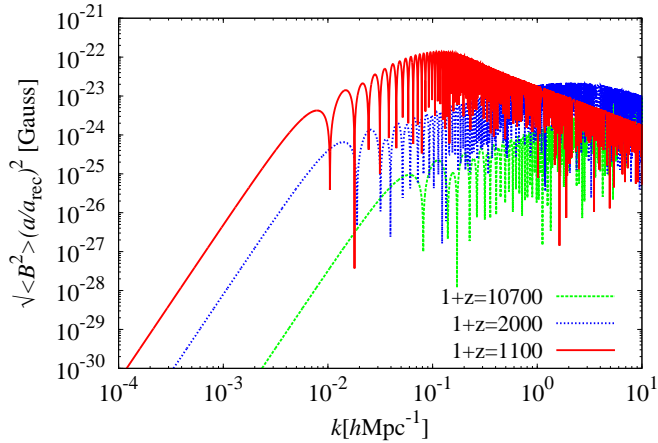}}
\caption{
The spectrum of magnetic fields with $c_{14}=-1.0\times 10^{-4}$, $c_{13}=-2.0\times 10^{-4}$, $\alpha=1.0\times 10^{-4}$, $c_{1}=-3.0\times 10^{-4}$ and $c^{2}_{v}=3.0$ at $1+z=1100$, $1+z=2000$ and $1+z=10700$.
The initial amplitude and the aether parameters are fixed as $\mathcal{A}_{V}c^{2}_{13}c^{-4}_{v}=2.0\times 10^{-17}$.
}
\label{fig:power1}
\end{figure}

\subsection{Power spectrum of magnetic fields}
The final results of generated magnetic fields spectra are depicted in Fig.~\ref{fig:power1}.
This figure shows that there are three characteristic scales. Let us
focus on the scales, $10^{-4}\lesssim k/h{\rm Mpc^{-1}}\lesssim 10^{-2}$,
which are outside the sound horizon at $1+z=2000$, right before the
recombination epoch. In these scales, the magnetic field spectrum is proportional to $k^{4}$.
Next, let us focus on the scales which are inside the sound horizon and in
the tight-coupling regime ($\delta v \ll v_\gamma$) , $10^{-2}\lesssim
k/h{\rm Mpc^{-1}}\lesssim 10^{0}$. In these scales the aether field begins
to oscillate after the sound horizon entry. The velocity difference
$\delta v$, the source of magnetic fields, also begins to oscillate
because $\delta v$ is dragged by the aether field through $\sigma$ (see
Eq.~(\ref{approx expression}). The velocity difference grows with
oscillations until the tight-coupling breaks down. In these scale, the spectrum of
magnetic fields is proportional to $k^{1}$ with oscillations.

Finally we focus on the scales, $k/h{\rm Mpc^{-1}}
\gtrsim 10^{0}$, in which the tight-coupling
approximation is no longer valid. In
these scales $\delta v \sim v_\gamma \sim \sigma$, and $\delta v$ begins
to decay as $\propto a^{-1}$ together with $\sigma$.
Consequently, the spectrum of magnetic fields decays with oscillations at $k/h{\rm
Mpc^{-1}}\gtrsim 10^{0}$. As a result the spectrum of magnetic fields becomes proportional to
$k^{-1}$ with oscillations. 

At $1+z=1100$, the amplitude of magnetic fields has an extra
enhancement due to the recombination process. When the Fourier mode is inside
the sound horizon and in the tight-coupling regime, baryons oscillate
with photons. The Fourier modes in this scale can have an extra growth
because the tight-coupling becomes weaker as the recombination process takes
place. On the other hand, there is little enhancement for the Fourier
modes at scales where the tight-coupling has already been broken down at recombination.
As a result, around the recombination epoch, the position of the peak of the spectrum moves
slightly toward
the smaller wavenumber.
\begin{figure}[htp]
\rotatebox{0}{\includegraphics[width=0.5\textwidth]{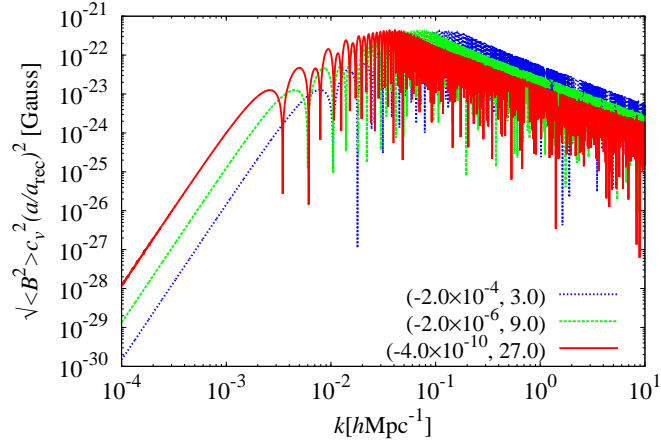}}
\caption{
The spectrum of magnetic fields with difference of the aether parameters and the initial amplitude at $1+z=1100$.
To normalize the amplitude of the spectra, we multiply by $c^{2}_{v}$.
We impose only the condition, this condition is same as Fig.~\ref{fig:CMB compare}, $\mathcal{A}_{V}c^{2}_{13}c^{-4}_{v}=2.0\times 10^{-17}$, on the aether parameters and the initial amplitude.
The spectra of the magnetic fields are depicted by solid lines,
$(c_{13},\ c^{2}_{v})=$
$(-2.0\times 10^{-4},\ 3.0)$,
$(-2.0\times 10^{-6},\ 9.0)$ and
$(-4.0\times 10^{-10},\ 27.0)$, respectively.
}
\label{fig:power compare}
\end{figure}

Let us now estimate the largest amplitude of the magnetic fields allowed
from the current observations. We find that the amplitude of the
generated magnetic fields also depends on the combination of ${\cal
A}_{V}c_{13}^{2}c^{-8}_{v}$. This aether parameter dependence can
be understood in the same way as CMB power spectra as follows. Here we
assume the universe is dominated by radiations.
Magnetic fields are given as the time integral of the velocity
difference $\delta 
v$ (Eq.(\ref{Mag})) as 
\begin{eqnarray}\begin{split}
\sqrt{a^{4}(t)\frac{k^{3}}{2\pi^{2}}S_{B}(k,t)}&=\left( \frac{4\sigma_{T}}{3e}\right)\sqrt{2\mathcal{P}_{V}(k)}k \int^{t}_{0}{dt' a(t')\rho_{\gamma}(t')\delta v(k,t')}\\
&\propto \int^{\eta}_{0}{d\eta' \eta'^{2}\rho_{\gamma}(\eta')\delta v(k,\eta')} \\
&\propto \int^{\eta}_{0}{d\eta' \eta'^{-2}\delta v(k,\eta')} \ .
\label{Mag2}
\end{split}\end{eqnarray}
Time dependence of $\delta v$ has two characteristic epochs; the sound 
horizon crossing and the tight-coupling breakdown. Before the sound
horizon crossing, $\delta v$ evolves as 
$\delta v \propto \eta^{6}$ as shown in Figs. \ref{fig:evolve12} and \ref{fig:evolve45}, (also see Eq.~~(\ref{approx expression})). Once
the Fourier mode enters the sound horizon, $\delta v$ evolves as $\delta
v \propto \eta^{4}$ with oscillations. When the
tight-coupling breaks down, moreover, $\delta v$ evolves as $\delta v \propto
\eta^{-1}$ with oscillations. 
Therefore, by substituting the above time dependence into
Eq.~(\ref{Mag2}), we obtain the time
dependence of magnetic fields as $B\propto \eta^5$, $\eta^2/c_v$,
and $\eta^{-3}/c_v$, before and after sound horizon crossing, and
after tight-coupling breakdown, respectively.
Note that the above dependence of $\eta^2/c_v$ and $\eta^{-3}/c_v$ are obtained
from the fact that $\int d\eta \eta^n e^{ikc_v\eta} \sim \eta^n/(i kc_v) 
e^{ikc_v\eta} \sim \eta^n/c_v$.
Accordingly, the dependence of magnetic fields on the aether parameters
can be understood in the same way as CMB power spectra.
$B\propto B_{\rm ini} (\eta_\ast/\eta_{\rm ini})^5 \left((\eta_{\rm tc}/\eta_*)^2/c_v\right)
\left((\eta/\eta_{\rm tc})^{-3}/c_v\right) \propto c_v^{-5} c_v^{-1} c_v^{2}
\propto c_v^{-4}$ where $\eta_\ast \propto c_v^{-1}$ and $\eta_{\rm tc}
\propto c_v^{-1}$ are the epochs
of sound horizon crossing and tight-coupling breakdown, respectively.
Since the amplitude of $\delta v$ is proportional to $c_{13}$, the overall
dependence is given by 
$B^2\propto {\cal A}_{V}c_{13}^{2}c^{-8}_{v}$, with ${\cal A}_{V}$ being
the initial power spectrum amplitude.
This dependence is explicitly depicted in Fig.~\ref{fig:power compare}.
If $\mathcal{A}_{V}c^{2}_{13}c^{-8}_{v}$ is fixed,
the amplitudes of generated magnetic fields are almost the same.
Note that, however, peak positions of the magnetic field spectrum
depend on $c_{v}$ as the CMB temperature anisotropy power
spectrum.

Let us estimate the maximum amount of magnetic fields allowed from the
limits of the CMB power spectra Eq.~(\ref{new constraint}) and the constraint on the sound speed $c_v$
 (Appendix A.5). 
We find that the spectrum of magnetic fields with aether field parameter
dependence is given by
\begin{eqnarray}
\sqrt{\Braket{B^{2}}}\sim
\left\{ \begin{array}{ll}
\displaystyle 10^{-19}\left( \frac{\mathcal{A}_{V} c_{13}^2c^{-4}_{v}}{2.0\times
	 10^{-17}} \right)^{1/2} \left(\frac{c_v^{2}}{3}\right)^{-1}\left( \frac{k}{0.01\; h{\rm Mpc^{-1}}}\right)^{4}\; \left[ {\rm Gauss} \right]&(k/h{\rm Mpc^{-1}} \lesssim 10^{-2}/c_{v})\\
\displaystyle 10^{-23}\left( \frac{\mathcal{A}_{V} c_{13}^2c^{-4}_{v}}{2.0\times 10^{-17}} \right)^{1/2}\left(\frac{c_v^{2}}{3}\right)^{-1}\left( \frac{k}{0.01\; h{\rm Mpc^{-1}}}\right)^{1}\; \left[ {\rm Gauss}\right]&(10^{-2}/c_{v}\lesssim k/h{\rm Mpc^{-1}} \lesssim 1/c_{v})\\
\displaystyle 10^{-21}\left( \frac{\mathcal{A}_{V} c_{13}^2c^{-4}_{v}}{2.0\times 10^{-17}} \right)^{1/2}\left(\frac{c_v^{2}}{3}\right)^{-1}\left( \frac{k}{0.01\; h{\rm Mpc^{-1}}}\right)^{-1}\; \left[ {\rm Gauss}\right]&(1/c_{v}\lesssim k/h{\rm Mpc^{-1}}) \ .\\
\end{array} \right. \label{mag spectrum}
\end{eqnarray}
If we set ${\cal A}_{V}c_{13}^{2}c^{-4}_{v}=2.0\times 10^{-17}$
(Eq.~(\ref{new constraint})) and
$c_{v}=1.0$, which satisfies the constraints in Appendix A, magnetic
fields have the largest amplitude as $10^{-22}$ Gauss at $0.1\; h{\rm
Mpc^{-1}}$.

\section{Summary}
In this paper, we explored a mechanism of the generation of
magnetic fields in the Einstein-Aether gravity. This theory contains a
dynamical vector field, i.e., the aether field, and the aether field
excites the metric perturbation of cosmological vector modes. Because
the metric perturbation generates the velocity difference between
baryons and photons, magnetic fields are naturally generated through
Thomson scattering.
This effect can arise only from the second order in the tight-coupling approximation.
We derived solutions analytically up to the second order in the tight-coupling approximation to understand the results of the numerical calculations.

We investigated evolution of vector perturbations in detail in
the Einstein Aether gravity. We found that the evolution of the vector metric
perturbation $\sigma$ is quite different from the one 
with vector perturbations in General Relativity~\cite{Ichiki:2011ah}. In
particular, in the case with 
the aether field the Silk damping does not arise even after the tight
coupling is broken down because $\sigma$ is
large enough to support the velocity perturbation of photons. 
We checked that the Silk damping arises again if we 
turn off $\sigma$ by hand.
We found that the amplitudes of CMB temperature and B-mode polarization power spectra depend on the initial
amplitude and the aether parameters as $C_{\ell}\propto
\mathcal{A}_{V}c^{2}_{13}c^{-4}_{v}$. By comparing the TT CMB power
spectrum induced from the aether field 
with latest observations, we obtained a new constraint on the aether
parameters as 
$\mathcal{A}_{V}c^{2}_{13}c^{-4}_{v} \lesssim 2.0\times 10^{-17}$.

In addition, we found the dependence of generated magnetic fields on the
initial amplitude and the aether parameters as
$\sqrt{\Braket{B^{2}}}\propto \mathcal{A}_{V}c^{2}_{13}c^{-8}_{v}$,
which is slightly different from $C_{\ell}$'s. We then estimated the
amplitude of magnetic fields with the aether parameters within the
limits of current observations. We found that the maximum amount of
magnetic fields can be as large as $\sqrt{\Braket{B^{2}}}\sim 10^{-22}$
Gauss at $k=0.1\; h{\rm Mpc^{-1}}$.

The shape of the magnetic field spectrum is also different from the one
with vector perturbations in General Relativity~\cite{Ichiki:2011ah}. First, the exponential
cut-off in the magnetic field spectrum found in~\cite{Ichiki:2011ah}
does not arise. The reason is exactly the same as in the case for the
velocity perturbation of photons mentioned above since the source term, i.e., velocities, can survive the Silk damping effect. Second, since the
vector-mode metric perturbation $\sigma$ decays inside the sound
horizon, the velocity difference $\delta v$ also decays together with $\sigma$, once
the tight-coupling between photons and baryons breaks down. Thus the
spectrum of magnetic fields has a characteristic peak near the Silk
damping scale at the recombination epoch. Moreover, around the 
recombination epoch, magnetic fields at larger scales than the peak scale
experience a little enhancement due to the breakdown of the tight
coupling, although the overall spectrum shape 
does not change very much.

It would be interesting to consider the generation of magnetic fields in
other modified gravity theories or other inflation models which can
amplify the initial power spectrum. For instance, the Einstein-Aether
theory can be extended to have a more general kinetic term, so called
$F(K)$ gravity~\cite{Zuntz:2010jp}. It may also be interesting to 
constrain the aether parameters from other cosmological
observations, for instance, the correlations or weak lensing effects of
large scale structure. We leave these subjects for future works.

\begin{acknowledgments}
This work was supported in part by a Grant-in-Aid for JSPS Research under Grant No.~22-7477 (MS), JSPS Grant-in-Aid for Scientific Research under Grants No.~24340048 (KI) and 22340056 (NS), and Grant-in-Aid for Nagoya University Global COE Program ``Quest for Fundamental Principles in the Universe: from Particles to the Solar System and the Cosmos'', from the Ministry of Education, Culture, Sports, Science and Technology~(MEXT) of Japan. This research has also been supported in part by World Premier International Research Center Initiative, MEXT, Japan.
\end{acknowledgments}

\appendix
\section{Parameter constraints}
In this appendix, we summarize the current observational and theoretical constraints on the aether parameters.

\subsection{Effective gravitational constant}
To have positive effective gravitational constants, $G_{\rm cos}$ and
$G_{\rm N}$, 
\begin{eqnarray}
\alpha &<&2 \\
c_{14} &>&-2 \ .
\end{eqnarray}

\subsection{BBN constraint}
The primordial helium abundance created by the Big-Bang Nucleosynthesis
(BBN) is affected through the expansion rate of the universe via the Friedmann equation.
Since the Friedmann equation (Eq.~(\ref{Einstein eq})) contains the aether
parameters, the measurements of the primordial helium abundance can
constrain the aether parameters as ~\cite{Carroll:2004ai}
\begin{equation}
c_{14}+\alpha \lesssim 0.2 \ .
\end{equation}
This inequality can be easily satisfied if we set $c_{14}=-\alpha$.
We assume this relation in our numerical calculations.

\subsection{PPN limits}
Parametrized post-Newtonian (PPN) parameters in the Einstein-Aether gravity
have been derived and constrained in refs.~\cite{Foster:2005dk,Eling:2003rd,Will:2005va}.
PPN parameters in the Einstein-Aether gravity are reduced to two parameters $\alpha_{1}$ and $\alpha_{2}$
which are related to the aether parameters $c_{1}, c_{2}, c_{3}$ and $c_{4}$.
The constraints from solar-system tests are written in the form as
\begin{eqnarray}\begin{split}
\alpha_{1}&\equiv \frac{-8(c^{2}_{3}+c_{1}c_{4})}{2c_{1}-c^{2}_{1}+c^{2}_{3}}\lesssim 1.7\times 10^{-4} \ ,\\
\alpha_{2}&\equiv \frac{\alpha_{1}}{2}-\frac{(2c_{13}-c_{14})(\alpha+c_{14})}{c_{123}(2-c_{14})}\lesssim 1.2\times 10^{-7} \ .
\end{split}\end{eqnarray}

\subsection{Stability of Scalar-Vector-Tensor perturbations}
The linear perturbations in the Einstein-Aether gravity have been analyzed~\cite{ArmendarizPicon:2010rs}.
Scalar-, vector- and tensor-perturbations must be stable in the context of quantum and classical treatments.
The classical stability imposes the condition that the square of the
sound speed must be positive.
The quantum stability imposes the condition that the ghost should not appear.
In other words, the coefficient of a kinetic term of each mode must be positive.
These conditions imply that
\begin{eqnarray}\begin{split}
&1+c_{13}>0 \ ,\\
&c_{14}\leq 0,\ c_{1}\leq \frac{c^{2}_{13}}{2(1+c_{13})} \ ,\\
&-2\leq c_{14}<0,\ c_{123}<0 \ .
\end{split}\end{eqnarray}

\subsection{Cherenkov radiation and Superluminal motion}
If the sound speed of metric perturbations is smaller
than the speed of light, the transverse-traceless graviton has
sub-luminal dispersion~\cite{Elliott:2005va}. Accordingly, 
relativistic particles will lose its energy
by emitting gravitons 
through a similar process of Cherenkov radiation.
Therefore we can constrain the aether parameters
from the fact that high energy cosmic rays have been observed on the
Earth.
However, we can avoid these constraints if we assume that all modes 
propagate super-luminally. The
conditions of superluminal propagation are given in 
ref.~\cite{ArmendarizPicon:2010rs} as
\begin{eqnarray}\begin{split}
(2+c_{14})c_{123}&\leq (2-\alpha)(1+c_{13})c_{14}\ ,\\
2c_{4}&\geq \frac{-c^{2}_{13}}{1+c_{13}}\ , \\
c_{13}&\leq 0 \ .
\end{split}\end{eqnarray}
The connection of a superluminal propagation and a violation of
causality is not trivial 
and has been still a matter of debate.
In this paper, we assume that all modes propagate superluminally for simplicity.
 
\subsection{Anisotropic stress}
Anisotropic stress of long-wavelength adiabatic modes in scalar
perturbations should not be 
too large~\cite{ArmendarizPicon:2010rs}. From this we have a constraint as
\begin{equation}
|c_{13}|\lesssim 1\ .
\end{equation}

\subsection{Isocurvature mode}
To prohibit isocurvature modes from being dominant at superhorizon scales, we find a constraint~\cite{ArmendarizPicon:2010rs}:
\begin{equation}
\frac{\alpha}{c_{14}}\geq -1~, \label{Iso}
\end{equation}
which is satisfied if we assume $c_{14}=-\alpha$.

\subsection{Radiation damping}
It is well known that the rate of orbital decay of the Hulse-Taylor
binary B1913+16 matches the one induced by the emission of gravitational
waves in General Relativity~\cite{Stairs:2003eg,Will:2001mx}.
In the Einstein-Aether gravity, the deviation of the rate from General
Relativity is controlled by the parameter ${\cal A}$
\begin{equation}
\mathcal{A}=\left( 1+\frac{c_{14}}{2}\right)\left[ \frac{1}{c_{t}}-\frac{2c_{14}c^{2}_{13}}{(2c_{1}+c^{2}_{1}-c^{2}_{3})^{2}}\frac{1}{c_{v}}-\frac{c_{14}}{6(2+c_{14})}\left( 3+\frac{2\alpha_{2}-\alpha_{1}}{2(2c_{13}-c_{14})}\right)^{2}\frac{1}{c_{s}}\right] \ ,
\end{equation}
where $\alpha_{1}$ and $\alpha_{2}$ are the PPN parameters, and
$c_{s,v,t}$ are the sound speed of scalar-, vector- and
tensor-perturbations~\cite{Foster:2006az}. To match the observed rate, the parameter $\mathcal{A}$ is constrained as
\begin{equation}
|\mathcal{A}-1|\lesssim O(10^{-3})\ .
\end{equation}

\section{Cosmological perturbations in Einstein-Aether gravity}
Recently, perturbed Einstein-Aether gravity is discussed and formulated by Refs.~\cite{Nakashima:2011fu,ArmendarizPicon:2010rs,Zuntz:2010jp,Lim:2004js,Zuntz:2008zz}.
Here we summarize the cosmological perturbation theory in the Einstein-Aether
gravity.
\subsection{Perturbed equations}
Here, we summarize the perturbed action and equations up to first order in real and Fourier spaces.
In the synchronous gauge, the metric is given by
\begin{equation}
ds^{2}=a^{2}\left[ -d\eta^{2}+\left( \delta_{ij}+h_{ij}\right)dx^{i}dx^{j}\right] \ .
\end{equation}
The aether field is written as 
\begin{eqnarray}\begin{split}
A^{\mu}&=\left( \frac{1}{a},\frac{V^{i}}{a}\right) \ ,\\
A_{\mu}&=\left( -a,aV_{i}\right) \ .
\end{split}\end{eqnarray}

First, the equation of motion for the aether field, i.e.,
Eq.~(\ref{Aether}) leads to following equations. The $\mu=0$ component reads
\begin{equation}
-a^{2}\delta\lambda =(c_{2}+c_{3})\dot{V}^{k}_{\ ,k}-(2c_{1}+c_{2}+c_{3})\mathcal{H}V^{k}_{\ ,k}+c_{2}\frac{1}{2}\ddot{h}^{k}_{\ k}-(2c_{1}+c_{2}+2c_{3})\frac{1}{2}\mathcal{H}\dot{h}^{k}_{\ k}~, \label{delta lambda}
\end{equation}
where $\delta \lambda$ is the variation of the Lagrange multiplier.
We use Eq.~(\ref{delta lambda}) to remove $\delta\lambda$ from perturbed
equations. The $\mu=i$ components are
\begin{eqnarray}\begin{split}
c_{14}\left( \ddot{V}_{i}+2\mathcal{H}\dot{V}_{i}\right)& -(\alpha -c_{14})\mathcal{\dot{H}}V_{i}+(\alpha +c_{14})\mathcal{H}^{2}V_{i} \\
&-\left[ c_{1}V_{i\ \ ,k}^{\ ,k}+(c_{2}+c_{3})V^{k}_{\ ,ik}+c_{2}\frac{1}{2}\dot{h}^{k}_{\ k,i}+(c_{1}+c_{3})\frac{1}{2}\dot{h}^{k}_{\ i,k}\right] =0 \ .
\end{split}\end{eqnarray}

Second, we rewrite the Einstein equation (\ref{Einstein}) as
\begin{eqnarray}\begin{split}
R_{\mu\nu}-\frac{1}{2}Rg_{\mu\nu}&=8\pi G\left( \frac{1}{8\pi G}T^{(A)}_{\mu\nu}+T^{(M)}_{\mu\nu}\right) \\
&\equiv 8\pi GT_{\mu\nu} \ .
\end{split}\end{eqnarray}
Then the perturbed Einstein equations lead to~\cite{Shiraishi:2012bh}:
\begin{eqnarray}\begin{split}
&\ddot{h}^{i}_{\ i}+\mathcal{H}\dot{h}^{i}_{\ i}=8\pi Ga^{2}\left( \delta T^{0}_{\ 0}-\delta T^{i}_{\ i}\right) \ ,\\
&\ddot{h}^{i}_{\ j}+2\mathcal{H}\dot{h}^{i}_{\ j}+\mathcal{H}\dot{h}^{k}_{\ k}\delta^{i}_{\ j}-\left( \partial^{i}\partial_{j}h^{k}_{\ k}+\partial^{k}\partial_{k}h^{i}_{\ j} -\partial^{k}\partial_{j}h^{i}_{\ k}-\partial_{k}\partial^{i}h^{k}_{\ j}\right)=16\pi Ga^{2}\left( \delta T^{i}_{\ j}-\frac{1}{2}\delta^{i}_{\ j}\delta T^{\mu}_{\ \mu}\right) \ , \\
&2\mathcal{H}\dot{h}^{i}_{\ i}+\partial^{j}\partial_{i}h^{i}_{\ j}-\partial^{k}\partial_{k}h^{i}_{\ i}=-16\pi Ga^{2}\delta T^{0}_{\ 0} \ ,\\
&\partial^{j}\dot{h}^{i}_{\ j}-\partial^{i}\dot{h}^{j}_{\ j}=16\pi Ga^{2}\delta T^{i}_{\ 0} \ .
\end{split}\end{eqnarray}
The perturbed energy momentum tensor for the aether field can be written in the form:
\begin{eqnarray}\begin{split}
a^{2}\delta T^{(A)}{}^{0}_{\ 0}=&\left[ c_{14}\dot{V}^{k}_{\ ,k}-(\alpha -c_{14})\mathcal{H}V^{k}_{\ ,k}-\alpha\frac{1}{2}\mathcal{H}\dot{h}^{k}_{\ k}\right] \ ,\\
a^{2}\delta T^{(A)}{}^{i}_{\ 0}=&-\left[ c_{14}\ddot{V}^{i}+2c_{14}\mathcal{H}\dot{V}^{i}-(\alpha -c_{14})\mathcal{\dot{H}}V^{i}\right. \\
&\left. +(\alpha +c_{14})\mathcal{H}^{2}V^{i}+\frac{1}{2}(c_{1}-c_{3})\left( V^{k,i}_{\ \ \ ,k}-V_{\ \ \ ,k}^{i,k}\right) \right] \ ,\\
a^{2}\delta T^{(A)}{}^{i}_{\ j}=&-\left[ c_{2}\left( \dot{V}^{k}_{\ ,k}+\frac{1}{2}\ddot{h}^{k}_{\ k}\right) +2c_{2}\left( \mathcal{H}V^{k}_{\ ,k}+\frac{1}{2}\mathcal{H}\dot{h}^{k}_{\ k}\right)\right]\delta^{i}_{\ j} \\
&-\left[ \frac{1}{2}c_{13}\left( \dot{V}^{i}_{\ ,j}+\dot{V}^{\ ,i}_{j}+\ddot{h}^{i}_{\ j}\right) +c_{13}\mathcal{H}\left( V^{i}_{\ ,j}+V^{\ ,i}_{j}+\dot{h}^{i}_{\ j}\right) \right] \ .
\end{split}\end{eqnarray}

We obey the convention of the Fourier transformation as
\begin{equation}
f({\bf x},\eta)=\int{\frac{d^{3}{\bf k}}{(2\pi)^{3}}f({\bf k},\eta){\rm e}^{i{\bf k\cdot x}}} \ .
\end{equation}
Furthermore we decompose these Fourier modes into scalar-, vector- and tensor-modes~\cite{Shiraishi:2012bh}.
The definitions of the scalar, vector and tensor decompositions are 
\begin{eqnarray}\begin{split}
&\omega^{i}({\bf k},\eta)=\omega^{(0)}\mathcal{O}^{(0)i}+\sum_{\lambda=\pm 1}\omega^{(\lambda)}\mathcal{O}^{(\lambda)i} \ ,\\
&\chi_{ij}({\bf k},\eta)=-\frac{1}{3}\chi_{\rm iso}\delta_{ij}+\chi^{(0)}\mathcal{O}^{(0)}_{ij}+\sum_{\lambda =\pm 1}\chi^{(\lambda)}\mathcal{O}^{(\lambda )}_{ij}+\sum_{\lambda =\pm 2}\chi^{(\lambda)}\mathcal{O}^{(\lambda)}_{ij} \ .
\end{split}\end{eqnarray}
where $\mathcal{O}{^{(\lambda)}_i}$ and $\mathcal{O}^{(\lambda)}_{ij}$
 are the projection operators for scalar
($\lambda=0$), vector ($\lambda=\pm 1$) and tensor ($\lambda=\pm 2$)
 modes. 
In the same way, the energy momentum tensor should be decomposed as
\begin{eqnarray}\begin{split}
&\delta T^{i}_{\ 0}({\bf k},\eta)=\delta T^{(0)}_{v}\mathcal{O}^{(0)i}+\sum_{\lambda =\pm 1}\delta T^{(\lambda)}_{v}\mathcal{O}^{(\lambda)i} \ ,\\
&\delta T^{i}_{\ j}({\bf k},\eta)=-\frac{1}{3}\delta T^{\rm iso}_{t}\delta^{i}_{\ j}+\delta T^{(0)}_{t}\mathcal{O}^{(0)}{}^{i}_{\ j}+\sum_{\lambda =\pm 1}\delta T^{(\lambda)}_{t}\mathcal{O}^{(\lambda)}{}^{i}_{\ j}+\sum_{\lambda =\pm 2}\delta T^{(\lambda)}_{t}\mathcal{O}^{(\lambda)}{}^{i}_{\ j} \ .
\end{split}\end{eqnarray}
The perturbed energy momentum tensor for ordinary matter is expressed as
\begin{eqnarray}\begin{split}
\delta T^{0}_{\ 0}&=-\rho \delta \ ,\\
\delta T^{\rm iso}_{t}&=-3p\Pi_{T} \ ,\\
\delta T^{(0)}_{v}&=-\left( \rho +p\right) v^{(s)}=-q^{(s)} \ ,\\
\delta T^{(0)}_{t}&=p\pi^{(s)} \ ,\\
\delta T^{(\pm 1)}_{v}&=-\left( \rho +p\right) v^{(v)}=-q^{(v)} \ ,\\
\delta T^{(\pm 1)}_{t}&=p\pi^{(v)} \ ,\\
\delta T^{(\pm 2)}_{t}&=p\pi^{(t)} \ .
\end{split}\end{eqnarray}

Then the Scalar-vector-tensor decomposition gives the following equations.
\begin{center}
\textit{Scalar-mode}
\end{center}
Equations of motion for the aether field are
\begin{eqnarray}\begin{split}
&a^{2}\delta\lambda =(c_{2}+c_{3})k\dot{V}^{(0)}-(2c_{1}+c_{2}+c_{3})\mathcal{H}kV^{(0)}+\frac{1}{2}c_{2}\ddot{h}_{\rm iso}-\frac{1}{2}(2c_{1}+c_{2}+2c_{3})\mathcal{H}\dot{h}_{\rm iso} \ , \\
&c_{14}\left( \ddot{V}^{(0)}+2\mathcal{H}\dot{V}^{(0)}\right) -(\alpha -c_{14})\mathcal{\dot{H}}V^{(0)}+(\alpha +c_{14})\mathcal{H}^{2}V^{(0)}+c_{123}k^{2}V^{(0)}+\frac{1}{3}c_{13}k\dot{h}^{(0)}+\frac{1}{6}\alpha k\dot{h}_{\rm iso}=0 \ .
\end{split}\end{eqnarray}
The scalar components of the energy momentum tensor for the aether field are
\begin{eqnarray}\begin{split}
a^{2}\delta T^{\rm iso}_{t}=&-\alpha \left[ k\dot{V}^{(0)}+2\mathcal{H}kV^{(0)}+\frac{1}{2}\ddot{h}_{\rm iso}+\mathcal{H}\dot{h}_{\rm iso}\right] \ , \\
a^{2}\delta T^{0}_{\ 0}=&-c_{14}k\dot{V}^{(0)}+(\alpha -c_{14})\mathcal{H}kV^{(0)}+\frac{1}{2}\alpha \mathcal{H}\dot{h}_{\rm iso} \ , \\
a^{2}\delta T^{(0)}_{v}=&-c_{14}\left( \ddot{V}^{(0)}+2\mathcal{H}\dot{V}^{(0)}\right) +(\alpha -c_{14})\mathcal{\dot{H}}V^{(0)}-(\alpha +c_{14})\mathcal{H}^{2}V^{(0)} \ , \\
a^{2}\delta T^{(0)}_{t}=&-\frac{1}{2}c_{13}\left( \ddot{h}^{(0)}+2\mathcal{H}\dot{h}^{(0)}\right) -c_{13}k\left( \dot{V}^{(0)}+2\mathcal{H}V^{(0)}\right) \ .
\end{split}\end{eqnarray} 
The Einstein equations are
\begin{eqnarray}\begin{split}
&\left( 1-\frac{\alpha}{2}\right)\left( \ddot{h}_{\rm iso}+\mathcal{H}\dot{h}_{\rm iso}\right) =(\alpha +c_{14})\left( k\dot{V}^{(0)}+\mathcal{H}kV^{(0)}\right) +8\pi Ga^{2}\left( \rho \delta +3p\Pi_{T}\right) \ , \\
&\left( 1-\frac{\alpha}{2}\right) \mathcal{H}\dot{h}_{\rm iso}+\frac{k^{2}}{3}\left( h_{\rm iso}-h^{(0)}\right) =-c_{14}k\dot{V}^{(0)}+(\alpha -c_{14})\mathcal{H}kV^{(0)}-8\pi Ga^{2}\rho \delta \ , \\
&k\left( \dot{h}_{\rm iso}-\dot{h}^{(0)}\right) =-3\left[ c_{14}\left( \ddot{V}^{(0)}+2\mathcal{H}\dot{V}^{(0)}\right) -(\alpha -c_{14})\mathcal{\dot{H}}V^{(0)}+(\alpha +c_{14})\mathcal{H}^{2}V^{(0)}\right]-24\pi Ga^{2}q^{(s)} \ , \\
&(1+c_{13})\left( \ddot{h}^{(0)}+2\mathcal{H}\dot{h}^{(0)}\right)+\frac{k^{2}}{3}\left( h_{\rm iso}-h^{(0)}\right) =-2c_{13}k\left( \dot{V}^{(0)}+2\mathcal{H}V^{(0)}\right) +16\pi Ga^{2}p\pi^{(s)} \ .
\end{split}\end{eqnarray}
\begin{center}
\textit{Vector-mode}
\end{center}
Equation of motion for the aether field is 
\begin{equation}
c_{14}\left( \ddot{V}^{(\pm 1)}+2\mathcal{H}\dot{V}^{(\pm 1)}\right) -(\alpha -c_{14})\mathcal{\dot{H}}V^{(\pm 1)}+(\alpha +c_{14})\mathcal{H}^{2}V^{(\pm 1)}+c_{1}k^{2}V^{(\pm 1)}+\frac{1}{2}c_{13}k\dot{h}^{(\pm 1)}=0 \ . \label{Vec aether}
\end{equation}
The vector components of the energy momentum tensor for the aether field are
\begin{eqnarray}\begin{split}
a^{2}\delta T^{(\pm 1)}_{v}=&-c_{14}\left( \ddot{V}^{(\pm 1)}+2\mathcal{H}\dot{V}^{(\pm 1)}\right)+(\alpha -c_{14})\mathcal{\dot{H}}V^{(\pm 1)}-(\alpha +c_{14})\mathcal{H}^{2}V^{(\pm 1)}-\frac{1}{2}(c_{1}-c_{3})k^{2}V^{(\pm 1)}\\
=&\frac{1}{2}c_{13}k^{2}V^{(\pm 1)}+\frac{1}{2}c_{13}k\dot{h}^{(\pm 1)}
		 \ , \label{EMTforAet}
\end{split}\end{eqnarray}
and
\begin{equation}
a^{2}\delta T^{(\pm 1)}_{t}=-\frac{1}{2}c_{13}\left[ k\dot{V}^{(\pm 1)}+2k\mathcal{H}V^{(\pm 1)}+\ddot{h}^{(\pm 1)}+2\mathcal{H}\dot{h}^{(\pm 1)} \right] \ ,
\end{equation}
where we used Eq.~(\ref{Vec aether}) to derive the second line from the first line in Eq.~(\ref{EMTforAet}).
The Einstein equations are
\begin{eqnarray}\begin{split}
\dot{\sigma}+2\mathcal{H}\sigma&=-\frac{c_{13}}{1+c_{13}}\left( \dot{V}+2\mathcal{H}V\right) -\frac{16\pi Ga^{2}p\pi^{(v)}}{(1+c_{13})k} \ , \\
k^{2}\sigma&=\frac{1}{1+c_{13}}\left[ 16\pi Ga^{2}q-c_{13}k^{2}V\right] \ ,
\end{split}\end{eqnarray}
where we defined new variable $\sigma\equiv \dot{h}^{(\pm 1)}/k$.
\begin{center}
\textit{Tensor-mode}
\end{center}
The tensor component of the the energy momentum tensor for the aether field is
\begin{equation}
a^{2}\delta T^{(\pm 2)}_{t}=-\frac{1}{2}c_{13}\left( \ddot{h}^{(\pm 2)}+2\mathcal{H}\dot{h}^{(\pm 2)}\right) \ .
\end{equation}
The Einstein equation is
\begin{equation}
\ddot{h}^{(\pm 2)}+2\mathcal{H}\dot{h}^{(\pm 2)}+\frac{k^{2}}{1+c_{13}}h^{(\pm 2)}=\frac{1}{1+c_{13}}16\pi Ga^{2}p\pi^{(t)} \ .
\end{equation}

\subsection{Perturbed action and initial power spectrum}
In this subsection, we calculate perturbed action up to second order to obtain the initial power spectrum of the aether field at the end of inflation.
For simplicity, the action is decomposed into each component as
\begin{equation}
S=S_{G}+S_{I}+S_{A} \ ,
\end{equation}
where
\begin{eqnarray}\begin{split}
&S_{G}= \frac{1}{16\pi G}\int{d^{4}x\sqrt{-g}R} \ ,\\
&S_{I}= \int{d^{4}x\sqrt{-g}\mathcal{L}_{I}} \ , \\
&S_{A}= \frac{1}{16\pi G}\int{d^{4}x\sqrt{-g}\mathcal{L}_{A}} \ .
\end{split}\end{eqnarray}
Here $\mathcal{L}_{I}$ is the Lagrangian density for the inflaton field.
Then the perturbed action up to second order is given by
\begin{eqnarray}\begin{split}
S^{(2)}_{G}=\frac{1}{16\pi G}\int{d^{4}x}\frac{a^{2}}{4}&\left[ \dot{h}^{k\ell}\dot{h}_{k\ell}-\dot{h}^{k}_{\ k}\dot{h}^{\ell}_{\ \ell}+\left( 4\mathcal{H}^{2}+5\mathcal{\dot{H}}\right) h^{k}_{\ k}h^{\ell}_{\ \ell}+\left( -2\mathcal{H}^{2}-4\mathcal{\dot{H}}\right)h^{k\ell}h_{k\ell} \right. \\
&\left. h^{k}_{\ k,j}h^{\ell \ ,j}_{\ \ell}-h^{k\ell}_{\ \ ,j}h^{\ \ ,j}_{k\ell}-2h^{k}_{\ k,\ell}h^{\ell j}_{\ \ ,j}+2h^{k}_{\ j,k}h^{\ell j}_{\ \ ,\ell}\right] \ ,
\end{split}\end{eqnarray}
\begin{equation}
S^{(2)}_{I}=\frac{1}{16\pi G}\int{d^{4}x}\frac{a^{2}}{4}\left[ \left( 2\mathcal{H}^{2}+4\mathcal{\dot{H}}\right)\left( h^{k\ell}h_{k\ell}-h^{k}_{\ k}h^{\ell}_{\ \ell}\right)-\alpha\left( \mathcal{H}^{2}+2\mathcal{\dot{H}}\right)\left( h^{k\ell}h_{k\ell}-h^{k}_{\ k}h^{\ell}_{\ \ell}\right)\right] \ ,
\end{equation}
\begin{eqnarray}\begin{split}
S^{(2)}_{A}=\frac{1}{16\pi G}\int{d^{4}x a^{2}}&\left[ -c_{14}\dot{V}^{i}\dot{V}_{i}+(\alpha +c_{14})\mathcal{H}^{2}V^{i}V_{i}-(\alpha -c_{14})\dot{\mathcal{H}}V^{i}V_{i} \right.\\
&+\frac{1}{4}\alpha \left( \mathcal{H}^{2}+2\mathcal{\dot{H}}\right) h^{k\ell}h_{k\ell}+\frac{1}{4}\alpha\left( \mathcal{H}^{2}-\mathcal{\dot{H}}\right)h^{k}_{\ k}h^{\ell}_{\ \ell} \\
&+c_{1}\left( V_{\ell}^{\ ,k}+\frac{1}{2}\dot{h}_{\ell}^{\ k}\right) \left( V^{\ell}_{\ ,k}+\frac{1}{2}\dot{h}^{\ell}_{\ k}\right) +c_{2}\left( V^{\ell}_{\ ,\ell}+\frac{1}{2}\dot{h}^{\ell}_{\ \ell}\right)\left( V^{k}_{\ ,k}+\frac{1}{2}\dot{h}^{k}_{\ k}\right) \\
&\left. +c_{3}\left( V^{k}_{\ ,\ell}+\frac{1}{2}\dot{h}^{k}_{\ \ell}\right) \left( V^{\ell}_{\ ,k}+\frac{1}{2}\dot{h}^{\ell}_{\ k}\right) \right] \ .
\end{split}\end{eqnarray}
Here, we employed the single field slow-roll inflation model and
replaced the inflaton field with ${\cal H}$ and metric perturbations by using the
Einstein equations and the equation of motion for the inflaton.

Hereafter we focus on the vector-mode only.
By performing Fourier transformation and scalar-vector-tensor decomposition, we have
\begin{equation}
S^{(2)}_{\rm vec}=\frac{1}{2}\int{d\eta}\int{\frac{d^{3}{\bf k}}{(2\pi)^{3}}}\left[ \left| \dot{v}\right|^{2}-\frac{\alpha}{c_{14}}\varepsilon \mathcal{H}^{2}\left| v\right|^{2}-c^{2}_{v}k^{2}\left| v\right|^{2}\right] \ ,
\end{equation}
where $v\equiv \sqrt{-c_{14}}aV/\sqrt{8\pi G}$ and $\mathcal{H}^{2}-\mathcal{\dot{H}}=\varepsilon \mathcal{H}^{2}$.
Through variation with respect to $v$, we have the equation of motion for the vector perturbation as
\begin{equation}
\ddot{v}_{k}+c^{2}_{v}k^{2}v_{k}+\frac{\alpha}{c_{14}}\varepsilon\mathcal{H}^{2}v_{k}=0 \ . \label{Vec quanta}
\end{equation}
Supposing that the slow-roll parameter $\varepsilon$ is constant during inflation, the conformal time $\eta$ and the scale factor $a$ are expressed as 
\begin{equation}
\eta =-\frac{1}{\mathcal{H}}\frac{1}{1-\varepsilon},\ \ \frac{a}{a_{I}}=\left( \frac{\eta}{\eta_{I}}\right)^{1/(\varepsilon-1)} \ ,
\end{equation}
where $\eta_{\rm I}$ and $a_{\rm I}$ are the conformal time and scale factor at the end of inflation.
Then we can solve Eq.~(\ref{Vec quanta}) easily to obtain
\begin{equation}
v_{k}(\eta)=\sqrt{\frac{\pi}{4}}(-\eta)^{1/2}{\rm e}^{i(2\nu_{\rm inf} +1)\pi /4}H^{(1)}_{\nu_{\rm inf}}(-kc_{v}\eta) \ ,
\end{equation}
where $\nu_{\rm inf}=\sqrt{\frac{1}{4}-\frac{\alpha}{c_{14}}\frac{\varepsilon}{(1-\varepsilon)^{2}}}$ and $H^{(1)}_{\nu}$ is the Hankel function of the first kind.
Finally the power spectrum of the aether field at the end of inflation is given by
\begin{equation}
\Braket{V(\eta,{\bf k})V^{*}(\eta,{\bf k'})}= (2\pi)^{3}\frac{2\pi^{2}}{k^{3}}\mathcal{P}_{V}(k)\delta({\bf k}-{\bf k'}) \ ,
\end{equation}
where
\begin{eqnarray}\begin{split}
&\mathcal{P}_{V}(k,\eta_{\rm I})=\frac{(1-\varepsilon)^{2}}{-c_{14}}\frac{H_{\rm I}^{2}}{(8\pi G)^{-1}}\Gamma^{2}(\nu_{\rm inf})\left( \frac{c_{v}}{2}\right)^{-2\nu_{\rm inf}}\left( -k_{0}\eta_{\rm I}\right)^{n_{v}}\left( \frac{k}{k_{0}}\right)^{n_{v}} \ ,
\end{split}\end{eqnarray}
with $k_{0}=0.002{\rm Mpc^{-1}}$ and $n_{v}=3-2\nu_{\rm inf}$.
From the above equation, we can read off the amplitude of the initial
power spectrum as 
\begin{equation}
\mathcal{A}_{V}=\frac{(1-\varepsilon)^{2}}{-c_{14}}\frac{H_{\rm
 I}^{2}}{(8\pi G)^{-1}}\Gamma^{2}(\nu_{\rm inf})\left(
						 \frac{c_{v}}{2}\right)^{-2\nu_{\rm
 inf}}\left( -k_{0}\eta_{\rm I}\right)^{n_{v}} \ . \label{ini amp}
\end{equation}
If we substitute the above amplitude into Eq.~(\ref{new constraint})
with the condition that $c_{14}=-\alpha$, the inequality Eq.~(\ref{new constraint})
can be rewritten as 
\begin{equation}
\frac{c^{2}_{13}}{(-c_{14}c^{5}_{v})}\lesssim 10^{43}\ , \label{renewconstraint}
\end{equation}
where we have assumed $\varepsilon \simeq 0.16$, which corresponds to
the scalar-tensor ratio 
$r \simeq 0.1$, $H_{\rm I} \simeq5\times
10^{13}$ GeV, and $\eta_{\rm I}\simeq - 2.0\times 10^{15}$ GeV.

%
%
\bibliography{ref}
\end{document}